\documentclass[draftclsnofoot,onecolumn,12pt]{IEEEtran}
\normalsize
\usepackage{amsmath}
\usepackage{amssymb}
\usepackage{bm}
\usepackage{mathtools}
\usepackage{cite}
\usepackage{amsthm}
\usepackage{graphicx}
\usepackage{comment}
\usepackage{tikz}
\usepackage{xfrac}
\usepackage{subcaption}
\usepackage{algpseudocode}
\usepackage{amsfonts}

\DeclareMathOperator\sign{sign}

\usepackage{graphicx}
\usepackage{float}
\usepackage{multirow}
\usepackage{tabularx}

\ifCLASSINFOpdf

\else

\fi

\newcommand\bcdot{\ensuremath{%
  \mathchoice%
   {\mskip\thinmuskip\lower0.2ex\hbox{\scalebox{1.2}{$\cdot$}}\mskip\thinmuskip}}%
   {\mskip\thinmuskip\lower0.2ex\hbox{\scalebox{1.2}{$\cdot$}}\mskip\thinmuskip}%
   {\lower0.3ex\hbox{\scalebox{1.2}{$\cdot$}}}%
   {\lower0.3ex\hbox{\scalebox{1.2}{$\cdot$}}}%
}
\begin{document}
\tikzstyle{decision} = [diamond, draw, text width=4.5em, text badly centered, node distance=3cm, inner sep=0pt]
\tikzstyle{block} = [rectangle, draw, text width=10em, text centered, rounded corners, minimum height=4em]
\tikzstyle{line} = [draw, -latex']
\tikzstyle{cloud} = [draw, ellipse, node distance=3cm, minimum height=2em]
\tikzstyle{red_block} = [draw, ellipse, fill = red!20, node distance=3cm, minimum height=2em]
\tikzstyle{dot} = []

\title{Analysis and Design of Serially Concatenated LDGM Codes}
\author{\IEEEauthorblockN{Amrit Kharel and Lei Cao}\\
\IEEEauthorblockA{Department of Electrical Engineering\\
University of Mississippi\\
University, MS 38677-1848\\
Email: amrit196@gmail.com and lcao@olemiss.edu}
}

\maketitle

\begin{abstract}
In this paper, we first present the asymptotic performance of serially concatenated low-density generator-matrix (SCLDGM) codes for binary input additive white Gaussian noise channels using discretized density evolution (DDE). We then provide a necessary condition for the successful decoding of these codes. The error-floor analysis along with the lower bound formulas for both LDGM and SCLDGM codes are also provided and verified. We further show that by concatenating inner LDGM codes with a high-rate outer LDPC code instead of concatenating two LDGM codes as in SCLDGM codes, good codes without error floors can be constructed. Finally, with an efficient DDE-based optimization approach that utilizes the necessary condition for the successful decoding, we construct optimized SCLDGM codes that approach the Shannon limit. The improved performance of our optimized SCLDGM codes is demonstrated through both asymptotic and simulation results.
\end{abstract}



\section{Introduction}
\begin{figure}
\centering{\includegraphics[width=4.5in]{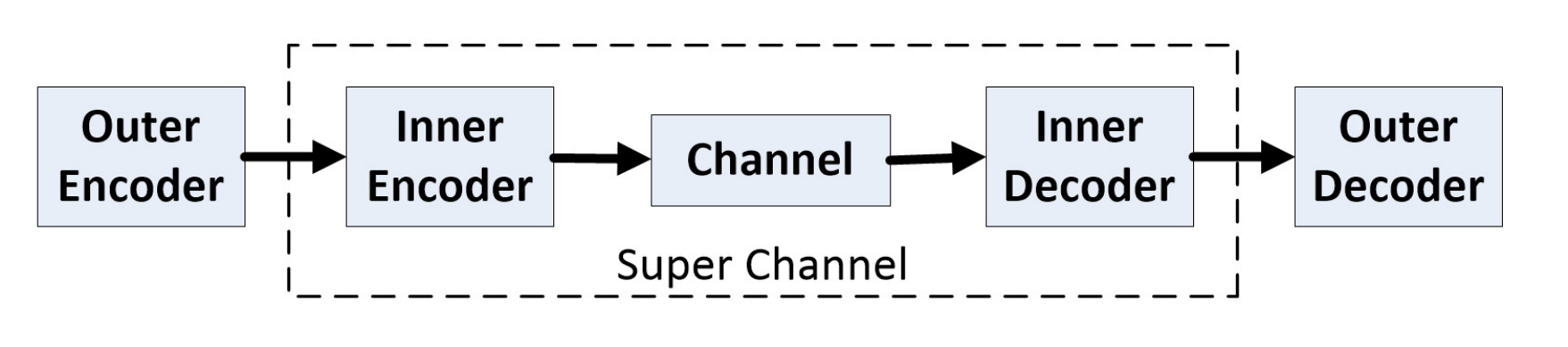}}
\caption{Concatenated Codes}
\label{fig_ConcatenatedCodes_BlockDiagram}
\end{figure}

Concatenated codes \cite{Concatened_Forney} were first introduced by David Forney in 1965. These codes employ two stages of encoding and decoding \cite{Concatened_Forney, book_ChannelCodes} as shown in Fig. \ref{fig_ConcatenatedCodes_BlockDiagram}. During encoding, the original information bits are first encoded using an outer code to produce intermediate bits which are further encoded using an inner code to produce final output bits that are transmitted over a communication channel. During decoding, first, the decoding of the inner code is conducted followed by the decoding of the outer code. Serially concatenated low-density generator-matrix (SCLDGM) codes that use the same two-stages of encoding and decoding are proposed in \cite{LDGM_original}. 
Although single LDGM codes are asymptotically bad \cite{LDGM_original}, the SCLDGM codes are shown to have impressive performance comparable to turbo and LDPC codes over noisy channels under sum-product (SP) decoding \cite{LDGM_original, LDGM_EXIT, LDGM_parallel}.

Both the inner and outer codes of the SCLDGM code use SP algorithm, also known as belief propagation (BP) algorithm, as the decoding method. The behaviour of the SP decoding can be exactly tracked using density evolution (DE) \cite{LDPC_DE_Richardson} or its discretized version called discretized density evolution (DDE) \cite{LDPC_ChungDissertation}. A DE approximation algorithm known as Gaussian approximation (GA) method \cite{LDPC_ChungDissertation} and extrinsic information transfer (EXIT) functions \cite{EXITchart} are also popularly used for analysing such codes that use SP decoding \cite{LDGM_parallel, LDGM_ErrorFloor, LDGM_EXIT, LDGM_LowRate}. In \cite{LDGM_original}, only simulation results are used to show that the SCLDGM codes perform close to the Shannon limit, however, no asymptotic analysis are presented for binary input additive white Gaussian noise (BIAWGN) channels. Asymptotic analysis of the single LDGM codes and the parallel concatenated LDGM codes are presented in \cite{LDGM_parallel}. In \cite{LDGM_ErrorFloor}, considering single LDGM codes two formulas are presented for the prediction of the error floor. EXIT functions are used in \cite{LDGM_EXIT, LDGM_ErrorFloor} to design better SCLDGM codes and provide their decoding thresholds. However, none of these works presented the exact asymptotic curves for the SCLDGM codes using the DDE. Since both the inner and outer decoders of SCLDGM codes use the SP decoding, it is possible to asymptotically analyse these codes using the DDE. Therefore, by applying the DDE to the inner code and then to the outer code, in this paper, we first provide the exact asymptotic curves and the decoding thresholds of the SCLDGM codes. Such DDE implementation for the concatenated codes termed as two-step DDE has been successfully developed and used in \cite{amritLetter2018, amritCCNC2018} to asymptotically analyse another class of concatenated codes known as physical layer Raptor codes \cite{RaptorSymmetric} and their systematic version.



Forney described \cite{Concatened_Forney} that the only function of the inner decoder is to bring the probability of decoding error to the range of $10^{-2} \sim 10^{-4}$, which once achieved, the outer decoder further drives the overall probability of decoding error down to an extremely small desired value. More specifically, that range for BIAWGN channel is predicted to be $10^{-2} \sim 10^{-3}$ \cite{Concatened_Forney}. Finding the exact value in that range is crucial for designing optimal concatenated codes. In this paper, for SCLDGM codes in BIAWGN channels that use the SP decoding algorithm, we provide the exact probability of decoding error, termed as critical bit error rate (BER), that an inner decoder should at least produce for the successful decoding of these codes. It also means that successful decoding is impossible if the decoding is handed to the outer decoder before the critical BER is achieved by the inner decoder. This necessary condition for successful decoding leads to a conclusion that in fact the overall decoding threshold of a concatenated code is the minimum $E_b/N_o$ at which the inner decoder achieves the critical BER determined by the outer code.

While LDGM codes are known to have poor decoding performance and bad error floors in BIAWGN channel \cite{LDGM_original}, the SCLDGM codes greatly improve the decoding performance. The error-floors are drastically lowered, however, are not eradicated. In order to understand the error-floor behaviour of these codes, in this paper, we present a detailed error-floor analysis and provide the lower-bound formulas for both the LDGM and SCLDGM codes. For the SCLDGM codes, we show that our lower-bound formulas become the closest approximation of the exact asymptotic performance in the region beyond their decoding thresholds. We also study the asymptotic performance of another concatenation scheme where inner LDGM codes are serially concatenated with high-rate regular low-density parity-check (LDPC) codes. This code was introduced in \cite{LDGM_LDPC} and termed as low-density parity-check and generator-matrix (LDPC-GM) code. It was shown that the LDPC-GM codes achieve capacity in binary erasure channel (BEC) under BP decoding. However, the exact asymptotic performance analysis of the LDPC-GM codes for BIAWGN channel that uses SP decoding using the DDE is still missing. Therefore, in this paper, we also provide the exact asymptotic curves of these codes using the two-step DDE method. Furthermore, we show that since the high-rate regular LDPC codes do not suffer from error-floors, its use as an outer code in the concatenated scheme can completely eradicate the error floors associated with LDGM codes, thus making the LDPC-GM codes error-floor-free capacity approaching codes for BIAWGN channels.


Density evolution based optimization approaches have been used to provide optimized LDPC codes with the best known decoding thresholds \cite{LDPC_ChungDissertation, LDPC_design_Richardson}. For the concatenated codes with a known outer code, due to the knowledge of the critical BER, designing a good concatenated code boils down to finding an inner code that achieves the critical BER at the lowest possible $E_b/N_o$ \cite{amritLetter2018}. This approach has been successfully used in \cite{amritLetter2018} to optimally design physical layer Raptor codes. In this paper, we further use the critical BER-based DDE optimization approach to design an optimized SCLDGM code that performs close to the Shannon limit. We demonstrate the improved performance of our optimized SCLDGM code through both asymptotic and simulation results.


The rest of the paper is organized as follows. Section \ref{section_EncoDeco} presents the encoding and decoding of the SCLDGM codes. In Section \ref{section_DDE_SCLDGM}, we provide the details of DDE implementation for SCLDGM codes to obtain the exact asymptotic curves and the decoding thresholds. We also provide the necessary condition for the successful decoding and discuss the faster convergence behaviour of these codes. Section \ref{section_ErrorFloorAnalysis} addresses the error-floor analysis of both the LDGM and SCLDGM codes along with the lower-bound formulas. In Section \ref{section_LDGM_LDPC}, we present the exact asymptotic curves of the LDPC-GM codes. We further show that by using high-rate LDPC code as an outer code, the error floors of concatenated codes are completely eradicated. Section \ref{section_Optimization} presents a capacity approaching optimized SCLDGM code designed using the critical BER-based DDE optimization approach. Finally, Section \ref{section_conclusion} concludes the paper.

\section{Encoding and Decoding of SCLDGM codes}\label{section_EncoDeco}
In the two-stage encoding of SCLDGM codes, at first, $k$ information bits are encoded using a high rate outer code to produce $k'$ intermediate bits which are further encoded using an inner code to produce $n$ output bits. Hence, the code-rates of the outer and the inner codes are $r_o=k/k'$ and $r_i=k'/n$ respectively, making a SCLDGM code of code-rate $r=r_i r_o = k/n$. Let $ \bm{s}=[s_1, s_2, \cdots, s_k]$ represent information bits. For a SCLDGM code, encoding at each stage is systematic. Therefore, the intermediate and output bits are $\bm {m}=[s_1, s_2, \cdots, s_k, t_1, \cdots, t_{k'-k}]$ and $\bm{o}=[s_1, s_2, \cdots, s_k, t_1, \cdots, t_{k'-k}, p_1, \cdots, p_{n-k'}]$ respectively, where $\bm{t}=[t_1, \cdots, t_{k'-k}]$ and $\bm{p}=[p_1, \cdots, p_{n-k'}]$ represent outer and inner parity bits. Let $n_i$ and $n_o$ represent the number of inner and outer parity bits respectively. Thus, $n=k+n_i+n_o$. These output bits are first modulated and then transmitted over a channel to be collected and decoded at the receiver.

\begin{figure}[ht!]
\centering
\begin{subfigure}[b]{0.5\textwidth}
\includegraphics[width=3.5in]{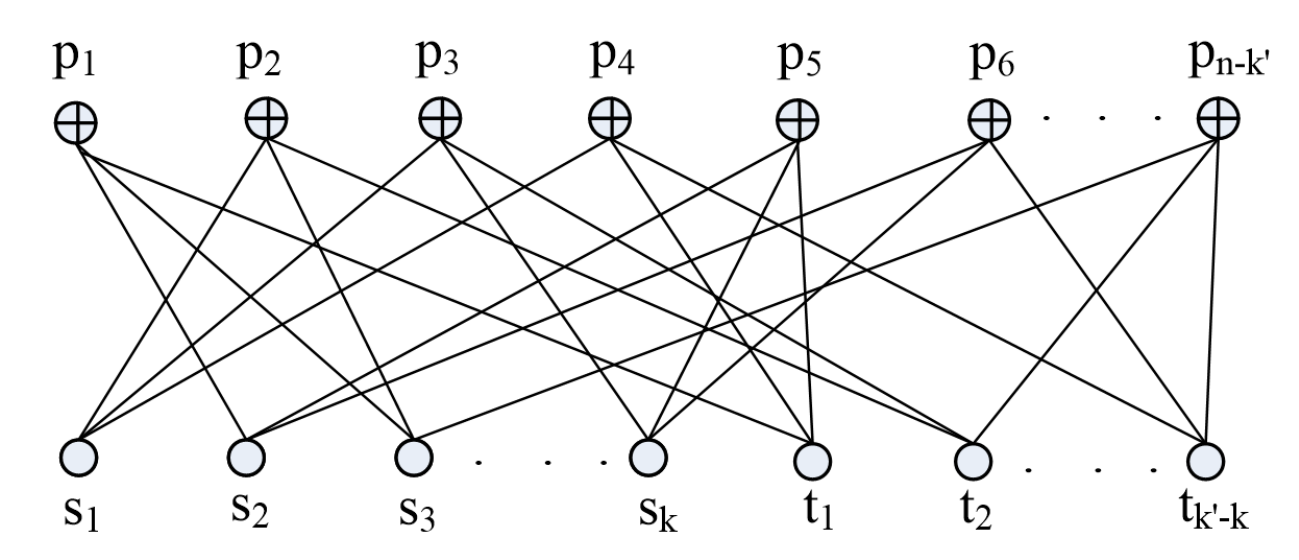}
\caption{Graph used for inner decoding}
\label{fig_innerDecodingGraph}
\end{subfigure}
\begin{subfigure}[b]{0.5\textwidth}
\includegraphics[width=3.5in]{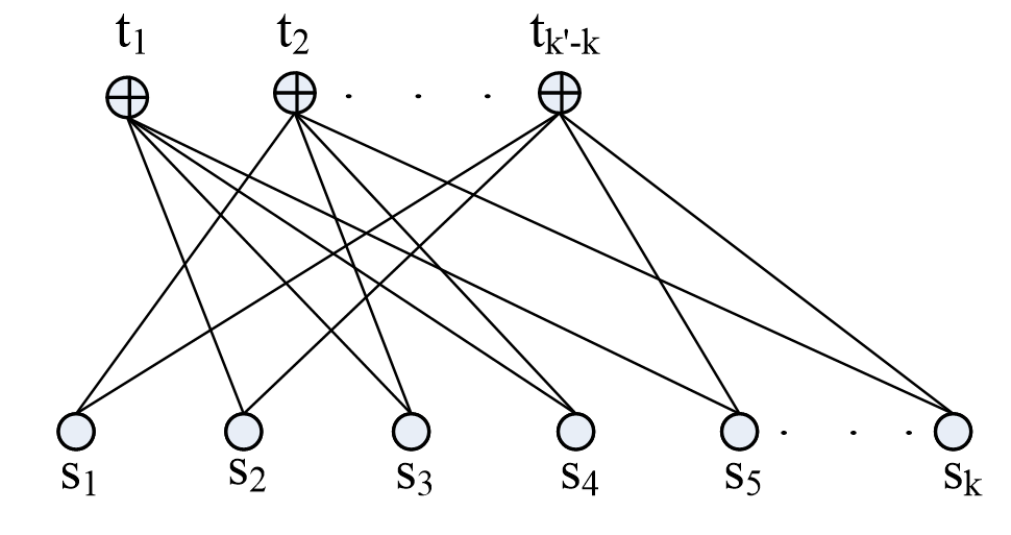}
\caption{Graph used for outer decoding}
\label{fig_outerDecodingGraph}
\end{subfigure}
\caption{An example of bi-partite graphs representing inner and outer decoders.}
\label{fig_DecodingGraphs}
\end{figure}

We consider transmission over the BIAWGN channel using binary phase shift keying (BPSK) modulation. This is modelled as $y=x+\eta$, where $x \in \{-1, +1\}$, $\eta$ is zero-mean Gaussian noise with $\sigma^2$ variance, and $y$ is the received corrupted bit. The channel estimate is calculated as $2y/\sigma^2$.

For the inner SP decoding, the bipartite graph as shown in Fig. \ref{fig_innerDecodingGraph} is used, where the inner parity bits $\left(\bm{p}=[p_1, \cdots, p_{n-k'}]\right)$ are used as check nodes (CNs) and intermediate bits $\left(\bm {m}=[s_1, s_2, \cdots, s_k, t_1, \cdots, t_{k'-k}]\right)$ are used as variable nodes (VNs). It must be noted that in this bipartite graph, both VNs and CNs are transmitted through the noisy channel and hence both have initial channel estimates. This is unlike the bi-partite graph used in LDPC codes where CNs do not have channel estimates.

Under LDGM's SP decoding, the updates from the $j$th CN to the $i$th VN denoted by $L_{c_jv_i}$ and from the $i$th VN to the $j$th CN denoted by $L_{v_ic_j}$ are computed as

\begin{equation}\label{eq_Lcjui}
{\small
\begin{split}
L_{c_jv_i} &= 2 \tanh ^{-1} \left (\left( \tanh\frac{L(c_j)}{2} \right) \prod_{i' \in N_c(j)- \{i\} } \tanh \left ( \frac{L_{v_{i'} c_j}}{2}\right ) \right)\\
L_{v_i c_j} &= L(v_i) + \sum_{j' \in N_v(i)- \{j\} } L_{c_{j'}v_i},
\end{split}
}
\end{equation}
where $L(v_i)$ and $L(c_j)$ represent the channel estimates of the corresponding VN and CN, respectively, and $N_c(j)$ denotes the set of VNs connected to the $j$th CN and $N_v(i)$ denotes the set of CNs connected to the $i$th VN.

The inner SP decoding is conducted for some predefined number of iterations and then the final estimate of each VN is calculated using a \textit{decision rule} $\left(L(v_i) + \sum_{j \in N_v(i) } L_{c_{j}v_i}\right)$ that sums all the incoming messages also known as log-likelihood ratios (LLRs) and the channel estimate of the VN. These values are then used as the initial LLRs for the outer decoding. The decoding graph used for outer decoding is as shown in Fig. \ref{fig_outerDecodingGraph}, where the outer parity bits are used as the CNs and the information bits are used as the VNs. After running SP decoding for the outer code over some predefined number of iterations, the final hard decision can be made for each VN based on the estimates calculated using the decision rule. We call such decoding where the decoding of the outer code is conducted only after completing the decoding of the inner code as two-step SP decoding.

\begin{figure}[ht!]
\centering{\includegraphics[width=5in]{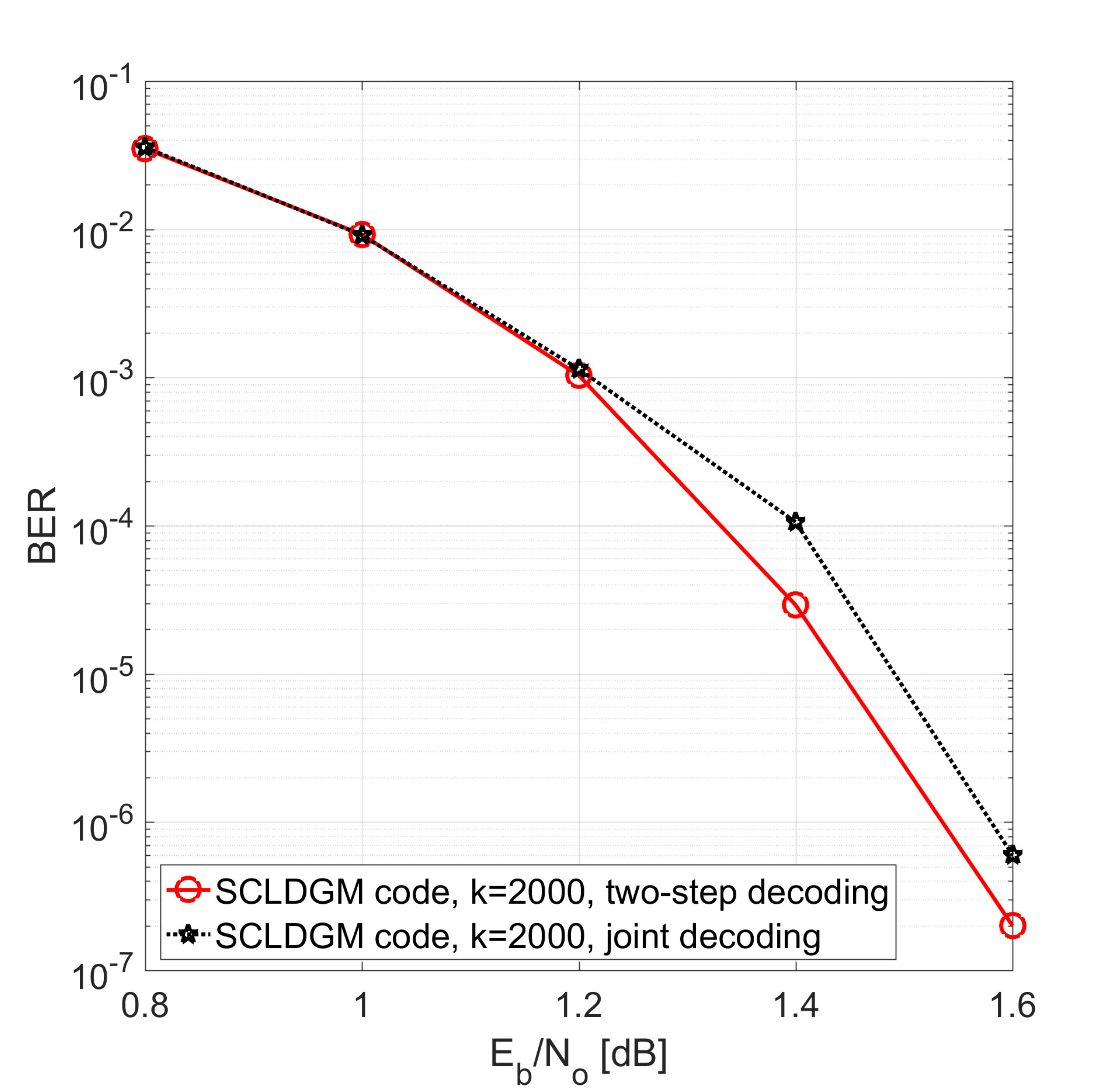}}
\caption{Two-step vs. joint decoding performance of a SCLDGM code. The inner code used is half-rate $(7,7)$-LDGM code. The outer code used is $50/51$-rate $(4,200)$-LDGM code. Overall code-rate = 0.49.}
\label{fig_TwoStepVsJoint_k2000}
\end{figure}

Decoding of such concatenated codes can also be done using joint decoding in which both the inner and outer decoding are conducted at each iteration and then the process is repeated over a predefined number of global iterations. Therefore, unlike in the two-step decoding where the outer decoder gets extra information at the end of the decoding of inner decoder, in the joint decoding, both the inner and outer decoder exchange information at each iteration. However, since the outer code considered in the SCLDGM codes is of very high rate, the information provided by it to the inner decoder is less significant. Thus, not much improvement in the decoding performance is achieved, which is also depicted in Fig. \ref{fig_TwoStepVsJoint_k2000}. We see that at the BER level of $10^{-5}$, the joint decoding merely outperforms the two-step decoding by a margin of around $0.05$ dB. Hence, for our asymptotic and error-floor analysis of the SCLDGM codes, we only consider the two-step SP decoding, which also follows the decoding method in \cite{LDGM_original, LDGM_EXIT}.

\section{Asymptotic Performance Analysis of SCLDGM codes}\label{section_DDE_SCLDGM}
It is known that the asymptotic behavior of the SP decoding can exactly be tracked using DE \cite{LDPC_DE_Richardson}. Both the inner and outer codes of SCLDGM codes use SP decoding. Hence, we first apply DDE to the inner code and then to the outer code to obtain the asymptotic curves of the SCLDGM codes. We assume that all-one BPSK symbols are transmitted. Therefore, the received LLRs (channel estimates) over a BIAWGN channel are known to be symmetric Gaussian distributed with mean $2/\sigma^2$ and variance $4/\sigma^2$, i.e., $\mathcal{N}\left(2/\sigma^2, 4/\sigma^2\right)$, where the variance is twice the mean and this symmetric condition has been proved to be preserved under the SP decoding \cite{LDPC_DE_Richardson}. The noise parameter $\sigma$, channel condition $E_b/N_o$, and the overall code-rate $r$ are related as $E_b/N_o=10 \log_{10}\left(1/(2r \sigma^2)\right)$.

\subsection{DDE for Inner Code}
\subsubsection{Degree Distributions}
For the inner LDGM code, the degree distributions of CNs and VNs from node perspective are given as $\Omega(x)=\sum_{i=1}^{k'}\Omega_i x^i$, where $\sum_{i=1}^{k'}\Omega_i=1$ and $\Lambda(x)=\sum_{i=1}^{n_i}\Lambda_i x^i$, where $\sum_{i=1}^{n_i}\Lambda_i=1$ respectively. Similarly, the degree distributions of CNs and VNs from the edge perspective are given as $\omega(x)=\sum_{i=1}^{k'}\omega_i x^{i-1}$, where $\sum_{i=1}^{k'}\omega_i = 1$ and $\lambda(x)=\sum_{i=1}^{n_i}\lambda_i x^{i-1}$, where $\sum_{i=1}^{n_i}\lambda_i=1$ respectively. The conversion between the node and edge degree distributions are done using $\omega_i = i\Omega_i / \sum_{j=1}^{k'}j\Omega_j$, and $\lambda_i = i\Lambda_i / \sum_{j=1}^{n_i}j\Lambda_j$ respectively.

\subsubsection{Implementation}
Let $\bar{v}$ be a quantized LLR message through a randomly chosen edge from a degree $d_v$ VN to a CN and let $\bar{u}$ be that from a degree $d_c$ CN to a VN. Under SP decoding, $\bar{v}$ is calculated as $\bar{v}=\sum_{i=0}^{d_v-1}\bar{u}_i,$ where $\bar{u}_i,\; i=1,\cdots, d_v-1$ are the incoming quantized LLRs from the neighboring CNs of the VN except the CN that gets the message $\bar{v}$, and $\bar{u}_0$ is the observed LLR of the VN.
As in \cite{LDPC_discretizedDE}, the probability mass function (PMF) of $\bar{v}$ is calculated as $p_{\bar{v}}=p_{\bar{u}_0}\bigotimes\left(\bigotimes^{d_v-1} p_{\bar{u}} \right)$, where $\bigotimes$ is discrete convolution while the superscript represents the number of times the convolution is operated, and $p_{\bar{u}}$ is the PMF of $\bar{u}$.
Similarly, for the discretized decision rule $\bar{D}=\sum_{i=0}^{d_v}\bar{u}_i$, the associated PMF can be calculated as $p_{\bar{D}}=p_{{\bar{u}}_0}\bigotimes\left(\bigotimes^{d_v} p_{\bar{u}} \right)$.

As in \cite{LDGM_ErrorFloor}, $\bar{u}$ is calculated as $\bar{u}=R(\bar{v}_0,R(\bar{v}_1, R(\bar{v}_2, \cdots, R(\bar{v}_{d_c-2},\bar{v}_{d_c-1}))))$, where ${v}_j,\; j=1,\cdots, d_c-1$ are the incoming LLRs from the neighboring VNs of the CN except the VN that gets the message $\bar{u}$, and $\bar{v}_o$ is the observed LLR of the CN. The two-input operation $R$ is defined as $R(a,b)=\mathbb{Q}\left( 2 \tanh^{-1}\left(\tanh\frac{a}{2}\tanh\frac{b}{2} \right)\right)$, where $a$ and $b$ are quantized messages, and $\mathbb{Q}$ is the quantization operator. The PMF $p_{\bar{u}}$ is computed as $ p_{\bar{u}} = R(p_{\bar{v}_0},R(p_{\bar{v}},R(p_{\bar{v}}, \cdots, R(p_{\bar{v}},p_{\bar{v}})))) = R(p_{\bar{v}_0},R^{d_c-1}p_{\bar{v}})$, where the PMF of $R(a,b)$ denoted as $p_c$ is computed as $p_c[k]=\sum_{(i,j):k\Delta = R(i\Delta,j\Delta)}p_a[i]p_b[j]$, where $\Delta$ is the quantization step.

By defining $f_{\lambda}(p)=\sum_{i=1}^{d_v}\lambda_i \left(p_{\bar{u}_0}\bigotimes \left( \bigotimes^{i-1}p\right) \right)$ and $f_{\omega}(p)=\sum_{j=1}^{d_c}\omega_j \left(R(p_{\bar{v}_0},R^{j-1}p\right)$, the evolving PMF of $\bar{u}$ and $\bar{v}$ at the $l$th iteration can then be calculated as $p_{\bar{v}}^{(l)} = f_{\lambda}\left(f_{\omega}\left(p_{\bar{v}}^{(l-1)}\right)\right)$ and $p_{\bar{u}}^{(l)}=f_{\omega}\left(f_{\lambda}\left(p_{\bar{u}}^{(l-1)}\right)\right)$ respectively, where the initial PMFs $p_{\bar{v}}^{(0)}$ and $p_{\bar{u}}^{(0)}$ have all mass at zero.

\subsubsection{Probability of Decoding Error for Inner Code}\label{section_Pe_LT}
For the inner code with all VNs having the same degree, the associated PMF of the decision rule at the $l$th iteration is $p^{(l)}_{\bar{D}}=p_{\bar{u}_0}\bigotimes\left(\bigotimes^{d_v} p_{\bar{u}} \right)$. However, for irregular inner code with the maximum VN degree of $d_{max}$, we can calculate $p^{(l)}_{\bar{D}}$ as

\begin{equation}\label{eq_pmfD1}
p_{\bar{D}}^{(l)}  =\sum_{i=1}^{d_{max}} \Lambda_i \left(p_{\bar{u}_0}\bigotimes \left(\bigotimes^{i}p_{\bar{u}}^{(l)}\right)\right).
\end{equation}

Let $[-L_a, L_a]$ be the range of LLRs used. Thus, the quantization step is calculated as $\Delta=2L_a/2^{n_b}$, where $n_b$ is the number of quantization bits used. Then, the decoding error probability of the inner decoder can be calculated at the completion of each iteration as
\begin{equation}\label{eq_Pe_LT_sum}
  E_{in}^{(l)}=\sum_{\bar{d}}p_{\bar{D}}^{(l)}(\bar{d}) \quad \text{for} \quad \bar{d} \in {[-L_a, -L_a+ \Delta, \cdots, 0]}.
\end{equation}


\subsection{DDE for Outer Code}
Let $q_{\bar{u}_{0}}$ be the PMF of the initial LLRs for the outer decoder. We know that after the completion of the inner decoding, the final LLRs of all the VNs of inner code are calculated using the decision rule and are then fed into the outer decoder. This means $p^{(l)}_{\bar{D}}$ serve as the PMF of the input LLRs to the outer decoder and is known to be $\mathcal{N}\left(M^{(l)}_{\bar{D}}, 2 M^{(l)}_{\bar{D}} \right)$, where $M^{(l)}_{\bar{D}}$ is the associated mean. Hence, we can write $q_{\bar{u}_{0}} = p_{\bar{D}}^{(l)}$. For the outer code, let $\Omega^{(o)}(x)$ and $\Lambda^{(o)}(x)$ be the degree distributions of the CNs and VNs from the node perspective, and $\omega^{(o)}(x)$ and $\lambda^{(o)}(x)$ be those from the edge perspective respectively. Let $q_{\bar{u}}^{(l)}$ be the PMF of the message through a randomly chosen edge from a CN to a VN at the $l$th outer decoding iteration. For the outer code with all VNs having the same degree $d_v^{(o)}$, the associated PMF of the decision variable at the $l$th outer iteration is $q^{(l)}_{\bar{D}}=q_{\bar{u}_0}\bigotimes\left(\bigotimes^{d_v^{(o)}} q_{\bar{u}} \right)$. However, for irregular outer code with the maximum VN degree of $d^{(o)}_{max}$, we can calculate $q^{(l)}_{\bar{D}}$ as
\begin{equation}\label{eq_pmfD2}
q^{(l)}_{\bar{D}} =\sum_{i=1}^{d^{(o)}_{max}} \Lambda^{(o)}_i \left(q_{\bar{u}_0}\bigotimes \left(\bigotimes^{i}q_{\bar{u}}^{(l)}\right)\right).
\end{equation}

Finally, the overall decoding error probability of the SCLDGM code is calculated as
\begin{equation}\label{eq_Pe_Raptor_sum}
  E^{(l)}=\sum_{\bar{d}}q_{\bar{D}}^{(l)}(\bar{d}) \quad \text{for} \quad \bar{d} \in {[-L_a, -L_a+ \Delta, \cdots, 0]}.
\end{equation}

Throughout this work, unless otherwise mentioned, we have used $n_b=10$, the range of LLRs used is $[-50, 50]$ and the maximum number of iterations used is 200 for all the asymptotic analysis.

\subsection{Necessary Condition for Successful Decoding}
In the two-stage decoding as shown in Fig. \ref{fig_ConcatenatedCodes_BlockDiagram}, we can consider the encoder-channel-decoder chain of the inner code as a super-channel \cite{Concatened_Forney}. Under DE, it is known that the input LLRs to the outer decoder has the PDF of $\mathcal{N}\left(M^{(l)}_D, 2 M^{(l)}_D \right)$. Therefore, the equivalent noise parameter, i.e., the noise standard deviation, associated with the super-channel assuming the unit signal power, can be calculated as $\sqrt{2/M^{(l)}_D}$. Let $\sigma^{(o)}_{th}$ be the decoding threshold of the outer code at and below which the outer code can theoretically achieve an extremely small BER. It is then clear that for the successful decoding, the outer decoder must satisfy the following condition
\begin{equation}\label{eq_conditionCriticalMD}
  \sqrt{\frac{2}{M^{(l)}_D}} \leq \sigma_{th}^{(0)}.
\end{equation}

Using the $\mathrm{Q}$-function, i.e., the upper tail function of the standard Gaussian distribution, (\ref{eq_conditionCriticalMD}) can be rewritten as
\begin{equation}\label{eq_conditionCriticalBER}
    \mathrm{Q}\left( \sqrt{\frac{M_D^{(l)}}{2}} \right) \leq \mathrm{Q}\left( \frac{1}{\sigma_{th}^{(o)}}\right).
\end{equation}

Under DE with all-one BPSK symbols transmission assumption, (\ref{eq_Pe_LT_sum}) can be rewritten as $E_{in}^{(l)} = \int_{-\infty}^{0}p_{D}^{(l)} dD$. Knowing that $p_D^{(l)} = \mathcal{N} \left(M_D^{(l)},2M_D^{(l)} \right)$, $E_{in}^{(l)}$ can be calculated as
\begin{equation}\label{eq_Pin}
  E_{in}^{(l)} = \int_{-\infty}^{0}p_{D}^{(l)} dD = 1 - \mathrm{Q}\left(\frac{0-M_D^{(l)}}{\sqrt{2M_D^{(l)}}} \right) = \mathrm{Q}\left( \sqrt{\frac{M_D^{(l)}}{2}} \right).
\end{equation}

From (\ref{eq_conditionCriticalBER}) and (\ref{eq_Pin}), we get
\begin{equation}\label{eq_conditionCriticalBERfinal}
  E^{(l)}_{in} \leq \mathrm{Q}\left( \frac{1}{\sigma_{th}^{(o)}}\right).
\end{equation}

In summary, (\ref{eq_conditionCriticalBERfinal}) shows the necessary condition that a SCLDGM code must satisfy in order to be successfully decoded in a BIAWGN channel using the SP algorithm. $\mathrm{Q}\left( 1/\sigma_{th}^{(o)}\right)$ is termed as critical BER. It is worth pointing out that in a BIAWGN channel under BPSK modulation, $\mathrm{Q}\left( 1/\sigma_{th}^{(o)}\right)$ is actually the raw input BER (i.e., the BER with direct threshold detection) of the outer code calculated at $\sigma_{th}^{(o)}$. Let this BER be represented by $P_{b_{th}}$. Then, (\ref{eq_conditionCriticalBERfinal}) can be rewritten as $E^{(l)}_{in} \leq P_{b_{th}}$,  which means that the raw input BER of the outer code at $\sigma_{th}^{(o)}$ is the critical BER. Clearly, handing the decoding to the outer decoder before this critical BER is achieved by the inner decoder can not produce successful decoding.

This necessary condition for the successful decoding asserts that achieving an extremely small BER by a SCLDGM code is equivalent to achieving the critical BER by the inner code. Hence, the decoding threshold of the overall SCLDGM code can be redefined as the minimum $E_b/N_o$ (maximum $\sigma$) at which the inner decoder achieves the critical BER that is determined solely by the outer code.
\subsection{Finding the Critical BER}
\begin{figure}[ht!]
\centering{\includegraphics[width=5.5in, height=3.5in]{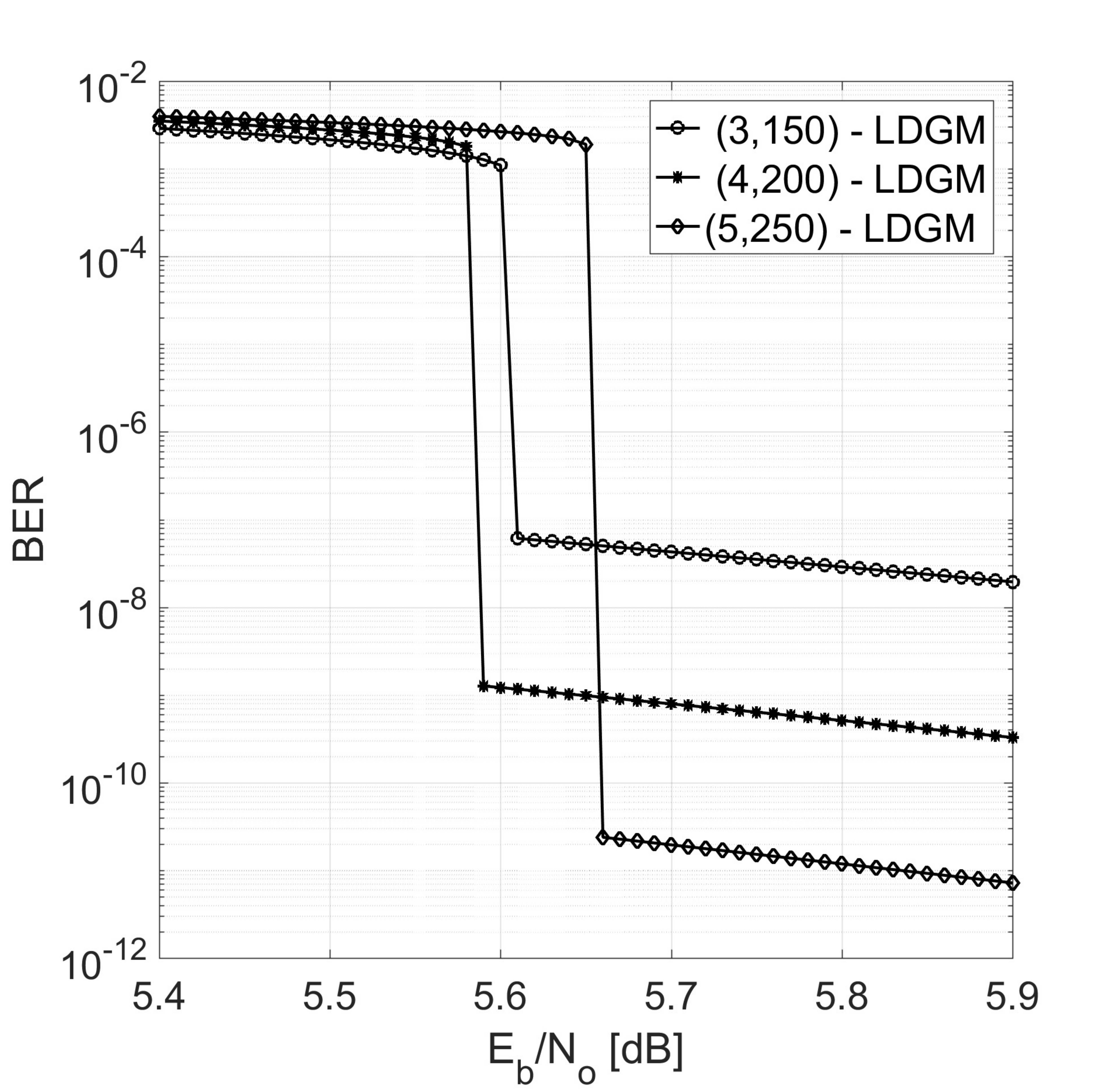}}
\caption{Asymptotic performance of $\left(d_v,d_c\right)$- LDGM codes, code-rate = 50/51.}
\label{fig_OuterCode50by51Performance}
\end{figure}

Fig. \ref{fig_OuterCode50by51Performance} presents the asymptotic performance obtained using DDE for three regular $(d_v,d_c)$-LDGM codes of code-rate $50/51$. Their decoding thresholds are presented in Table \ref{table_thresholds50by51}. Therefore, if these high rate LDGM codes are to be used as the outer codes in a SCLDGM code, the inner decoder must at least produce in each case the critical BER presented in Table \ref{table_thresholds50by51} in order to secure an overall successful decoding. Fig. \ref{fig_OuterCode50by51Performance} also shows that the $(3, 150)$ code is inferior to the $(4, 200)$ code in both error floor and decoding threshold, while the $(4, 200)$ code has better decoding threshold with a slightly higher error floor than the $(5, 250)$ code. We pick the $(4, 200)$ code as the outer code to further study the performance of SCLDGM codes. The outer code-rate $50/51$ is chosen since it is known to be one of the best choices for such serially concatenated codes \cite{LDGM_EXIT}. 

\begin{table}[ht]
\centering
\caption{DDE threshold of $50/51$-rate LDGM code and the corresponding critical BER.}
\label{table_thresholds50by51}
\scalebox{1}{
\begin{tabular}{|c| c| c| c|}
\hline
$\left(d_v, d_c \right)$ & $(E_b/N_o)_{th}$ & $\sigma^{(o)}_{th}$ & $\mathrm{Q}\left(1/\sigma^{(o)}_{th}\right)$ \\
\hline
$\left(3,150\right)$ & 5.61 & 0.374 & 3.778 $\times 10^{-3}$ \\ 
$\left(4,200\right)$ & 5.59 & 0.375 & 3.848 $\times 10^{-3}$ \\ 
$\left(5,250\right)$ & 5.66 & 0.372 & 3.608 $\times 10^{-3}$ \\ 
\hline
\end{tabular}
}
\end{table}

\subsection{Asymptotic Curves and Decoding Thresholds}
\begin{figure}[ht!]
\centering{\includegraphics[width=5.5in, height=4.0in]{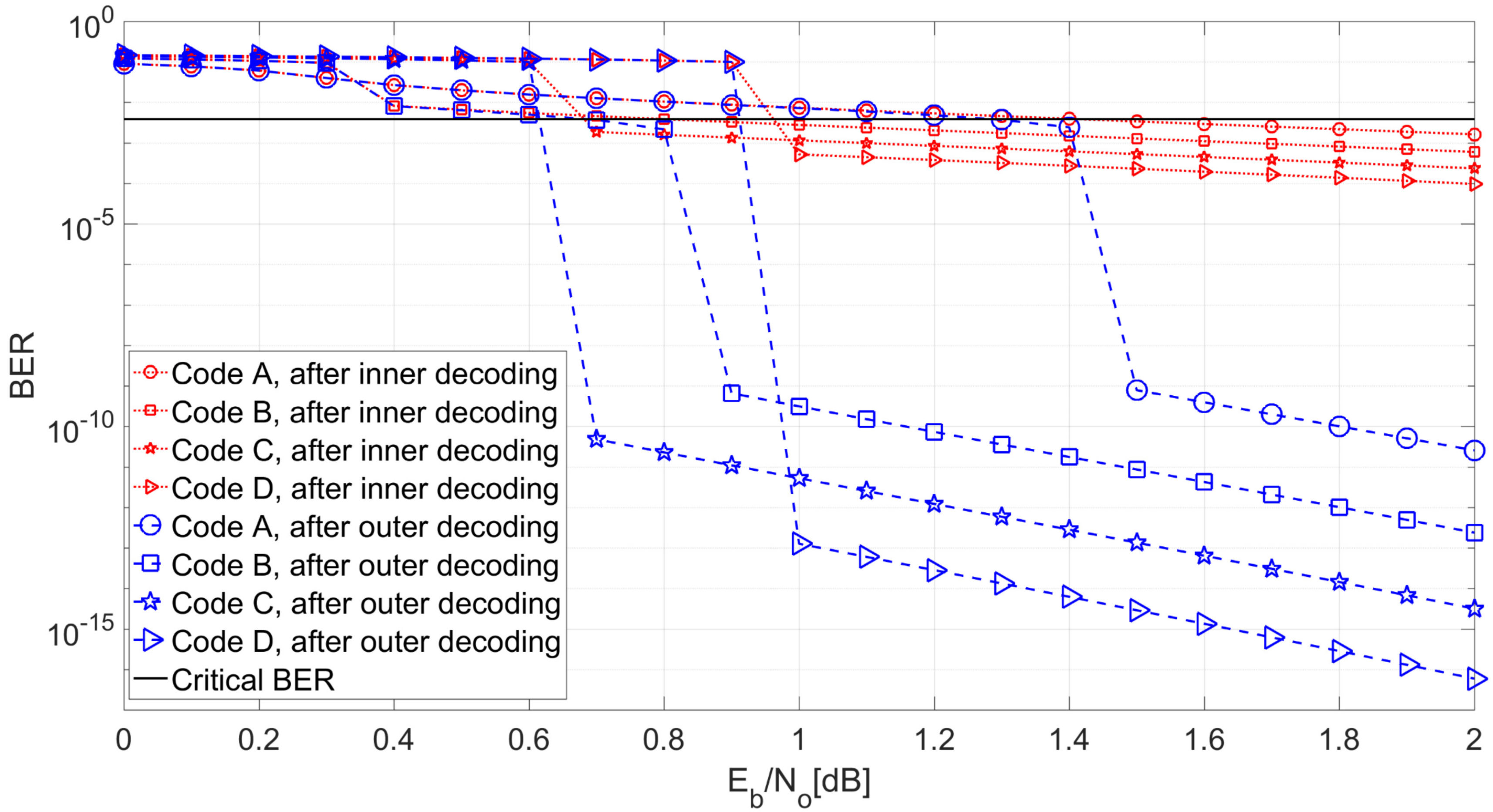}}
\caption{Asymptotic performance of various SCLDGM codes. Overall code-rate = 0.49.}
\label{fig_XXldgm_criticalBER}
\end{figure}

The asymptotic performance obtained for four SCLDGM codes, namely code A, B, C, and D, are presented in Fig. \ref{fig_XXldgm_criticalBER}. The inner codes used for code A, B, C, and D are the regular $\left(5,5\right)$, $\left(6,6\right)$, $\left(7,7\right)$, and $\left(8,8\right)$ half-rate LDGM codes respectively while the outer code used in each case is the $50/51$-rate (4,200)-LDGM code. Hence, the overall code-rate is $r=25/51$. These curves are obtained using the two-step DDE method detailed earlier. The curves clearly depict that once the inner decoder produces a BER below the critical BER, the outer decoder in each case drastically reduces the BER to the level enough to declare successful decoding. For example, for code A, the BERs computed after inner decoding from $0$ to $1.4$ dB are above the critical BER, i.e., above $3.848\times 10^{-3}$. The BERs further computed after outer decoding in that range mostly remain the same and only show some sign of improvement near $1.4$ dB. However, at $1.5$ dB and beyond, the BERs computed after inner decoding start falling below the critical BER, which after outer decoding immediately fall below $10^{-9}$ level. We observe such facts holding true in the case of each code, which validate the necessary condition for the successful decoding presented earlier. These curves may also justify the use of two-step decoding since the outer code has insignificant effect before the necessary condition is met by the inner code. We have also presented the decoding thresholds and the gap to the Shannon limit of these SCLDGM codes in Table \ref{table_SCLDGMthreshold}. We see that the Code C has the best decoding threshold of $0.68$ dB which is about $0.53$ dB away from the Shannon limit.

\begin{table}[ht]
\centering
\caption{DDE thresholds of various SCLDGM Codes. Overall code-rate = 0.49.}
\label{table_SCLDGMthreshold}
\scalebox{1}{
\begin{tabular}{|c| c| c| c|}
\hline
Code  & Threshold (dB) & Gap (dB) \\
\hline
\text{Code A} = (5,5)-LDGM + (4,200)-LDGM & 1.44 & 1.29 \\
\text{Code B} = (6,6)-LDGM + (4,200)-LDGM & 0.82 & 0.67 \\
\text{Code C} = (7,7)-LDGM + (4,200)-LDGM & 0.68 & 0.53 \\
\text{Code D} = (8,8)-LDGM + (4,200)-LDGM & 0.99 & 0.84\\
\hline
\end{tabular}
}
\end{table}

\subsection{Convergence of the Inner Decoder}
\begin{figure}[ht!]
\centering{\includegraphics[width=5.5in]{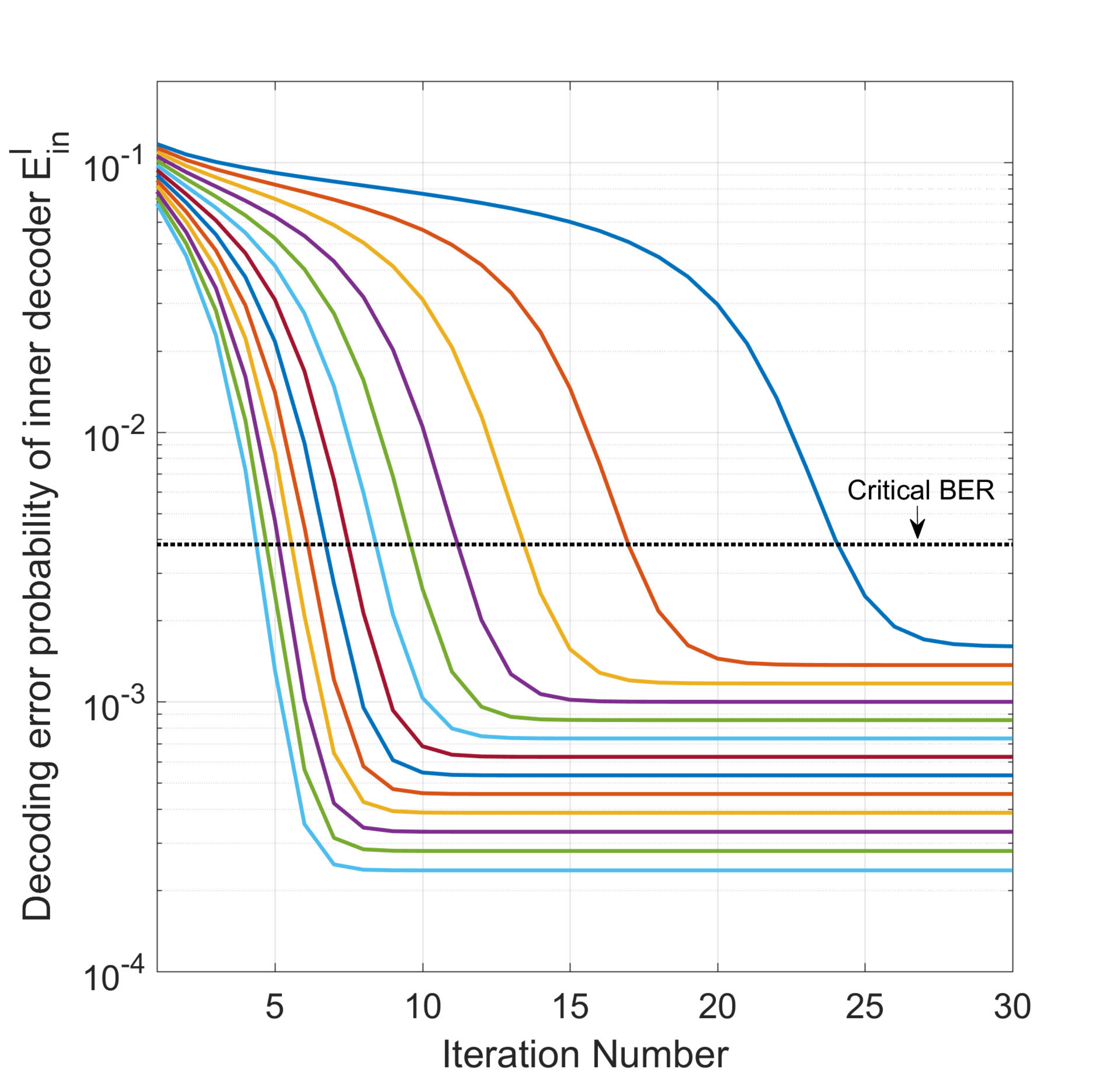}}
\caption{Decoding error probability of the inner decoder $\left(E_{in}^{(l)}\right)$ vs. iteration number for a SCLDGM code (Code C) at different $E_b/N_o$ values. From right to left in the graph, $E_b/N_o$ used is 0.8 to 2.0 dB with a step of 0.1 dB.}
\label{fig_BERinnerDecoderVsItrNum}
\end{figure}

Since the outer code used in SCLDGM code is of very high rate, the total number of edges in its decoding graph is very small compared to that of low-rate inner code. In addition, it is known that once the critical BER is achieved by the inner decoder, the number of iterations required by the outer decoder to provide successful decoding is very small. Due to these reasons, the decoding complexity added by the outer decoder is not much. On the other hand, due to the low code-rate of inner code that results in the larger decoding graph, the most of the decoding complexities of SCLDGM codes are associated with the number of iterations required by the inner LDGM codes to achieve the critical BER. Practically, the faster the inner decoding converges below the critical BER, the smaller is the decoding complexity. The asymptotic curves plotted in Fig. \ref{fig_BERinnerDecoderVsItrNum} shows that the inner decoder of the SCLDGM codes converges very fast, i.e., the number of iterations required by the inner decoder to provide BER well below the critical BER is small. Hence, this faster convergence behaviour makes the SCLDGM codes practically suitable codes capable of providing lower latency.

\section{Error-Floor Analysis of LDGM and SCLDGM codes}\label{section_ErrorFloorAnalysis}

Single LDGM codes are asymptotically bad and suffer from high error floors. The SCLDGM codes drastically improve the decoding performance and have decoding performance close to the Shannon limit. However, they still exhibit error-floor behaviour, although at the very lower BER level as shown in Fig. \ref{fig_XXldgm_criticalBER}. In this section, we provide the reasons behind such error floors and give the lower bounds for both the single LDGM and SCLDGM codes. For this purpose, we have considered both the inner and outer LDGM codes as regular codes with VN degree $d_v$ and $d_v^{(o)}$ respectively.

\subsection{Lower Bounds for single LDGM Codes}\label{section_LowerBoundSingleLDGM}
The message $u$ passing through a randomly chosen edge from a degree $d_c$ CN to a VN is calculated using the product rule as given by (\ref{eq_Lcjui}), which can be rewritten as
\begin{equation}\label{eq_ProdRuleRedef}
  u=2 \tanh^{-1} \prod_{j=0}^{d_c-1}\tanh(v_j/2),
\end{equation}
where $v_o$ is the channel estimate of the CN and $v_j, j=1,2,\cdots,d_c-1$ are the incoming LLRs from the neighboring VNs.

\begin{figure}[ht!]
\centering{\includegraphics[width=5.5in, height=4.2in]{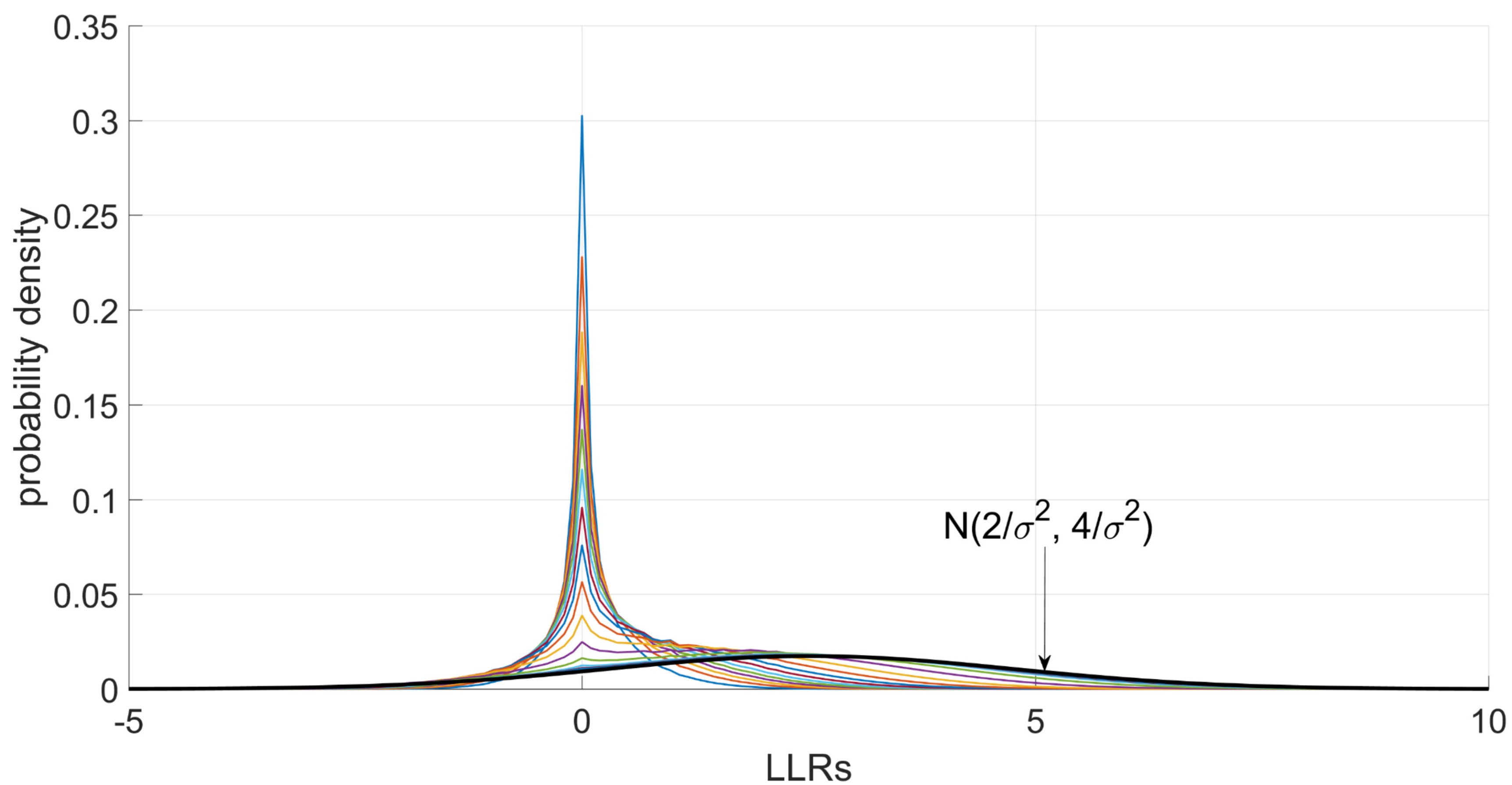}}
\caption{Evolution of $p_u^{(l)}$ towards right as $l$ increases for a $(7,7)$-LDGM code, $E_b/N_o$ = 1dB.}
\label{fig_Pu_itrAll_SNR1dB}
\end{figure}

\begin{figure}[ht!]
\centering{\includegraphics[width=5.5in, height=4.2in]{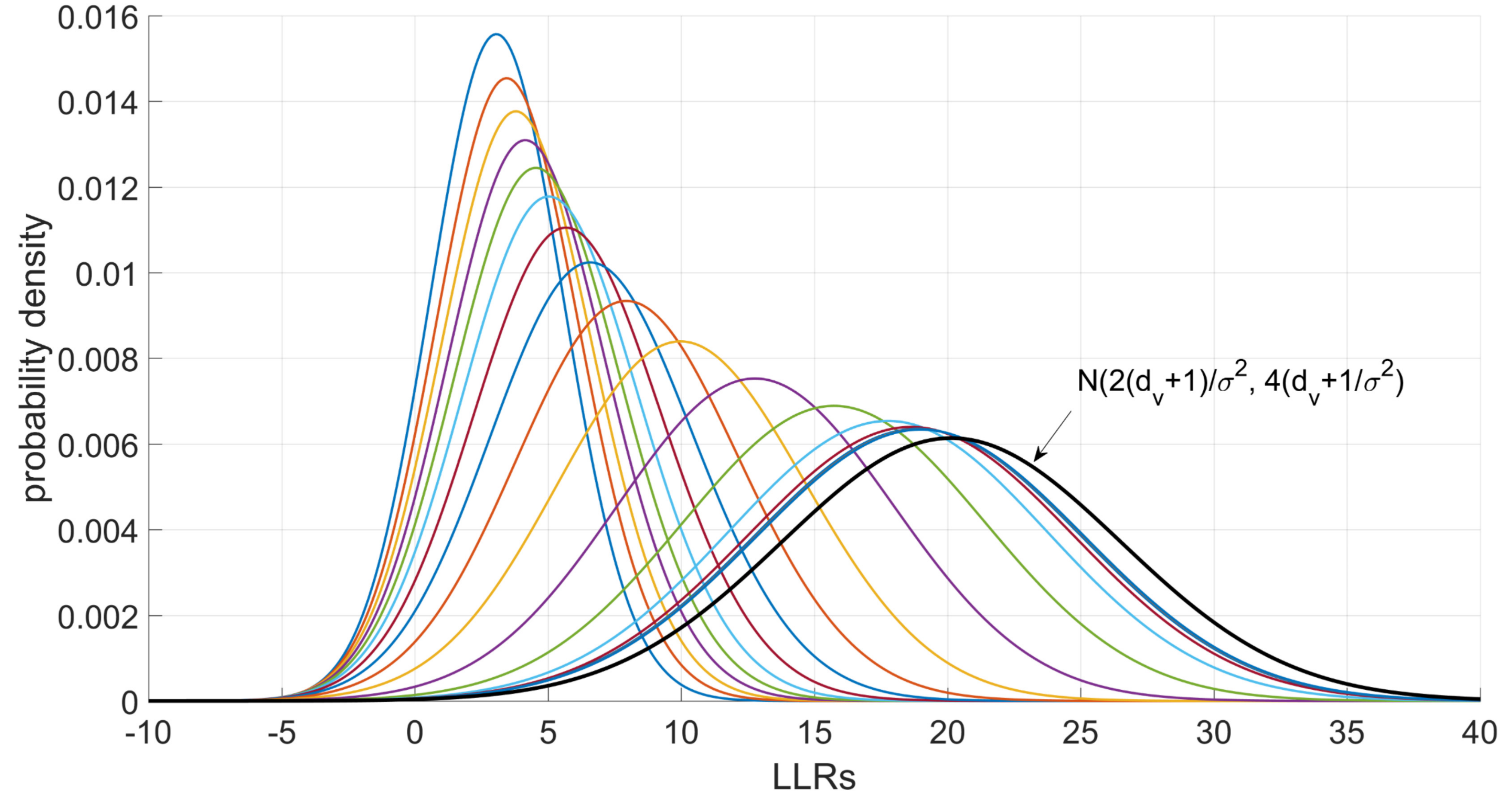}}
\caption{Evolution of $p_D^{(l)}$ towards right as $(l)$ increases for a $(7,7)$-LDGM code, $E_b/N_o$ = 1dB.}
\label{fig_PD_itrAll_SNR1dB}
\end{figure}

It is well known that (\ref{eq_ProdRuleRedef}) can be approximated by
\begin{equation}\label{eq_MinSum_u}
u \approx \left ( \prod_{j=0}^{d_c-1} \sign(v_j)\right )\cdot \min |v_j|.
\end{equation}
In fact, considering the magnitude of the LLR messages, we can prove the following result.

\vspace{0.08in}
{\bf Lemma} 1: For $u$ and $v_i, i=0,1,\dots, d_v-1$, defined in (10), we have
\begin{equation}\label{eq_uLessthan}
  |u| \leq \min \left( |v_0|, |v_1|, \cdots, |v_{d_c-1}| \right).
\end{equation}

{\it Proof}: presented in the Appendix.
\vspace{0.08in}

It is important to note that during the iterative decoding of LDGM codes, the messages $v_j, \; j=1, \cdots, d_c-1$ can continuously evolve, however, the message $v_0$ which is the channel estimate remains the same throughout the decoding process. Let $v_m = \min \left( |v_1|, |v_2|, \cdots, |v_{d_c-1}| \right)$. Hence, $|u| \leq \min \left( |v_0|, v_m  \right)$. Therefore, no matter how large the incoming LLRs are, due to the fact that $|v_0|$ remains the same, the following condition holds true throughout the decoding process
\begin{equation}\label{eq_uLessthanV0}
|u| \leq |v_o|.
\end{equation}

Hence, the magnitude of the LLR message being passed by a CN of an LDGM-based code is always upper bounded by the magnitude of its channel estimate.

It is known that $p_{v_0} \sim \mathcal{N}\left(2/\sigma^2, 4/\sigma^2\right)$. Under DE, the message $u$ is also assumed to be Gaussian distributed and consistent throughout the decoding iterations, i.e., $p_u^{(l)} \sim \mathcal{N}\left(M_u^{(l)}, 2 M_u^{(l)} \right)$, where $M_u^{(l)}$ is the mean associated with $u$ at the $l$th iteration.
In terms of the mean value, (\ref{eq_uLessthanV0}) can be written as
\begin{equation}\label{eq_MuLessthanMv0}
  M_{|u|}^{(l)} \leq M_{|v_o|} \Rightarrow M_{u}^{(l)} \leq M_{v_o}
\end{equation}

We can calculate the decoding error probability of $u$ as $e_u^{(l)} = Q\left( \sqrt{M_{u}^{(l)}/2} \right)$. Using (\ref{eq_MuLessthanMv0}), we obtain the following inequality
\begin{equation}\label{eq_ber_u_inequality}
  e_u^{(l)} \geq Q\left( \sqrt{\frac{M_{v_0}}{2}} \right)
\end{equation}

Therefore, even if $l \rightarrow \infty$, the decoding error probability of the message being passed by a CN is always lower bounded by $Q\left( \sqrt{M_{v_0}/2} \right)$. Knowing from (\ref{eq_MuLessthanMv0}) that the best possible $M_u^{(l)}$ during the DE is $M_{v_0}$, it is clear that the $p_u^{(l)}$ can never evolve beyond $\mathcal{N}\left(2/\sigma^2, 4/\sigma^2\right)$. This fact is clearly depicted in Fig. \ref{fig_Pu_itrAll_SNR1dB}, where $p_u^{(l)}$ for $l=1\; \text{to}\; 50$ are plotted along with $p_{v_0}$. We see that $p_u$ gradually evolves towards right as iteration increases, however, never evolves beyond $p_{v_0}$.

We know that the decision of each VN is made using the decision rule $D=\sum_{i=0}^{d_v}u_i = u_0 + \sum_{i=1}^{d_v}u_i$. From DE, we know that $u_1, u_2, \cdots, u_{d_v}$ are i.i.d Gaussian random variables. Hence, the mean associated with the $D$ at the $l$th iteration is $M_D^{(l)} = M_{u_0} + d_v M_u^{(l)}$. Since, $M_{u_0} = M_{v_0}$ and $M_u^{(l)} \leq M_{v_0}$, we can write the following
\begin{equation}\label{eq_MD_inequality}
  M_D^{(l)} \leq \left(d_v+1\right) M_{u_0}.
\end{equation}

\begin{figure}[ht!]
\centering{\includegraphics[width=5.5in]{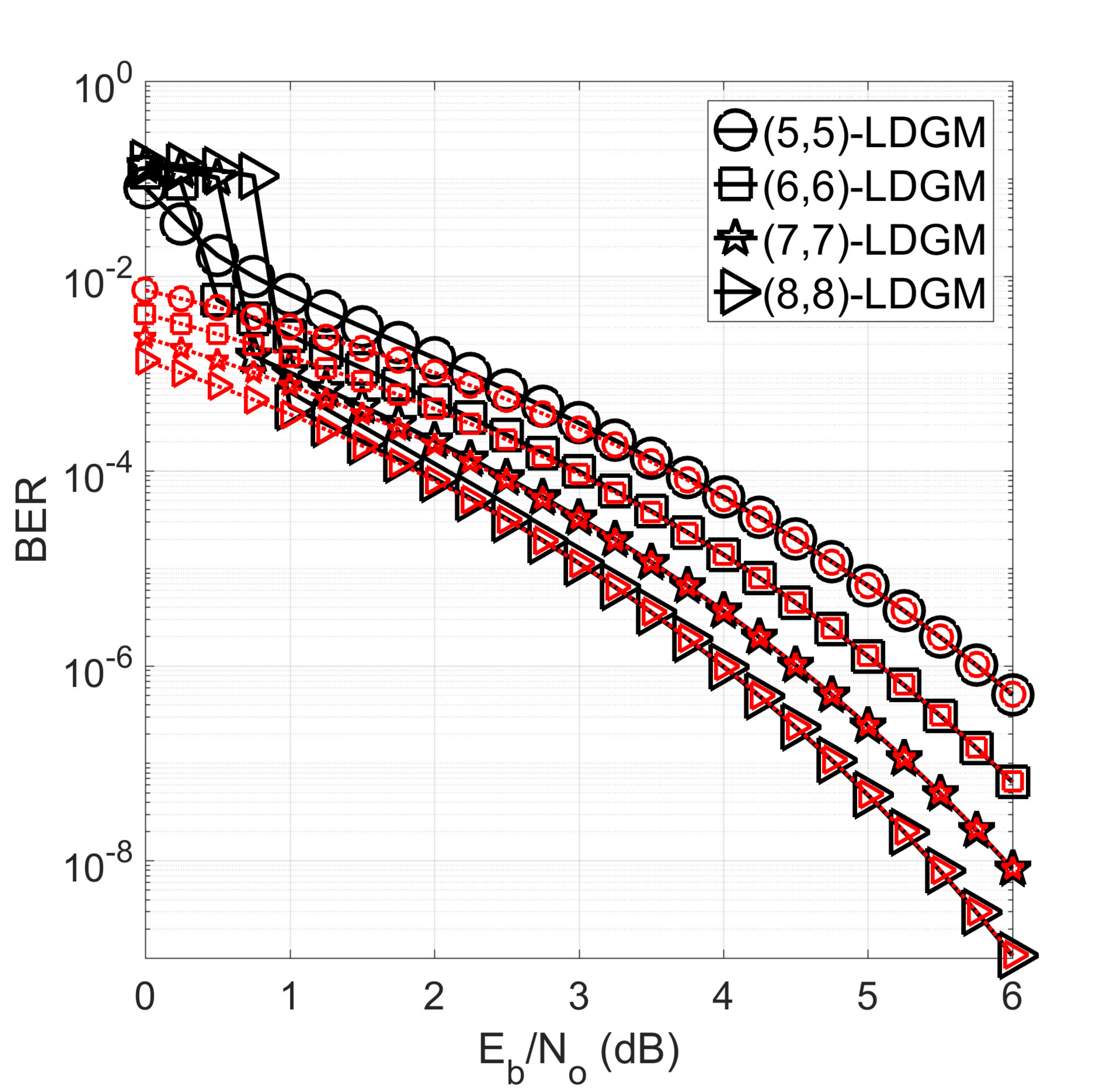}}
\caption{Asymptotic performance (solid lines) and lower bounds (dotted lines) of $\left(d_v,d_c\right)$-regular half-rate LDGM codes.}
\label{fig_DDE_LDGMhalfRate_LowerBound}
\end{figure}

Knowing that $p^{(l)}_{D} \sim \mathcal{N}\left(M^{(l)}_D, 2 M^{(l)}_D \right)$ and $M_D^{(l)} \leq \left(d_v+1\right) M_{u_0}$, it is clear that $p^{(l)}_{D}$ can never evolve beyond $\mathcal{N}\left(2(d_v+1)/\sigma^2, 4(d_v+1)/\sigma^2\right)$ which is clearly depicted in Fig. \ref{fig_PD_itrAll_SNR1dB}, where $p_D^{(l)}$ for $l=1\; \text{to}\; 50$ are plotted. As can be seen, initially the $p_D^{(l)}$ gradually evolves towards right as iteration increases, however, is bounded by $\mathcal{N}\left(2(d_v+1)/\sigma^2, 4(d_v+1)/\sigma^2\right)$.

The decoding error probability of $D$ calculated at the $l$th iteration is known to be $e_D^{(l)}=Q\left( \sqrt{M_{D}^{(l)}/2} \right)$. Using (\ref{eq_MD_inequality}), we obtain the following inequality
\begin{equation}\label{eq_eD_inequality}
\begin{split}
  e_D^{(l)} & \geq Q\left( \sqrt{\frac{(d_v+1)M_{u_0}}{2}} \right) \\
  & \geq Q\left( \sqrt{\frac{(d_v+1)}{\sigma^2}} \right)
\end{split}
\end{equation}

Hence, the decoding error probability of $D$, i.e., the BER of the LDGM codes is lower bounded by $Q\left( \sqrt{(d_v+1) / \sigma^2} \right)$. In Fig. \ref{fig_DDE_LDGMhalfRate_LowerBound}, the exact asymptotic performance of various regular half-rate LDGM codes obtained using the DDE is presented along with their respective lower-bounds calculated using $Q\left( \sqrt{(d_v+1)/\sigma^2} \right)$. We observe that at higher $E_b/N_o$ values, the exact BER of LDGM codes is closely bounded by the lower bound formula, i.e., $e_D \approxeq Q\left( \sqrt{(d_v+1) / \sigma^2} \right)$. Significant discrepancy occurs at lower $E_b/N_o$ values, which is due to the fact that some LLR values passing in the bipartite graph are even smaller than the direct channel estimate and hence dominate the performance.

\subsection{Lower Bounds for SCLDGM Codes}
Since the outer code of a SCLDGM code is also an LDGM code, the messages that the CNs of the outer decoder pass are also bounded by their respective initial estimates, and hence also suffer from error floors as we already observed in Fig. \ref{fig_XXldgm_criticalBER}.

It is known from the DE analysis of the SCLDGM codes under two-step decoding that the PDF of the input LLRs to the outer decoder is Gaussian with variance twice the mean. Let $\sigma_{sup}$ be the noise parameter associated with the super-channel. Hence, the input LLRs to the outer decoder has the PDF of $\mathcal{N}\left(2/\sigma_{sup}^2, 4/\sigma_{sup}^2\right)$. Hence, based on the analysis from Section \ref{section_LowerBoundSingleLDGM}, we can write the following
\begin{equation}\label{eq_E_bound1}
  E^{(l)} \geq Q\left( \sqrt{\frac{(d_v^{(o)}+1)}{\sigma_{sup}^2}} \right)
\end{equation}

\begin{figure}[ht!]
\centering{\includegraphics[width=5.5in]{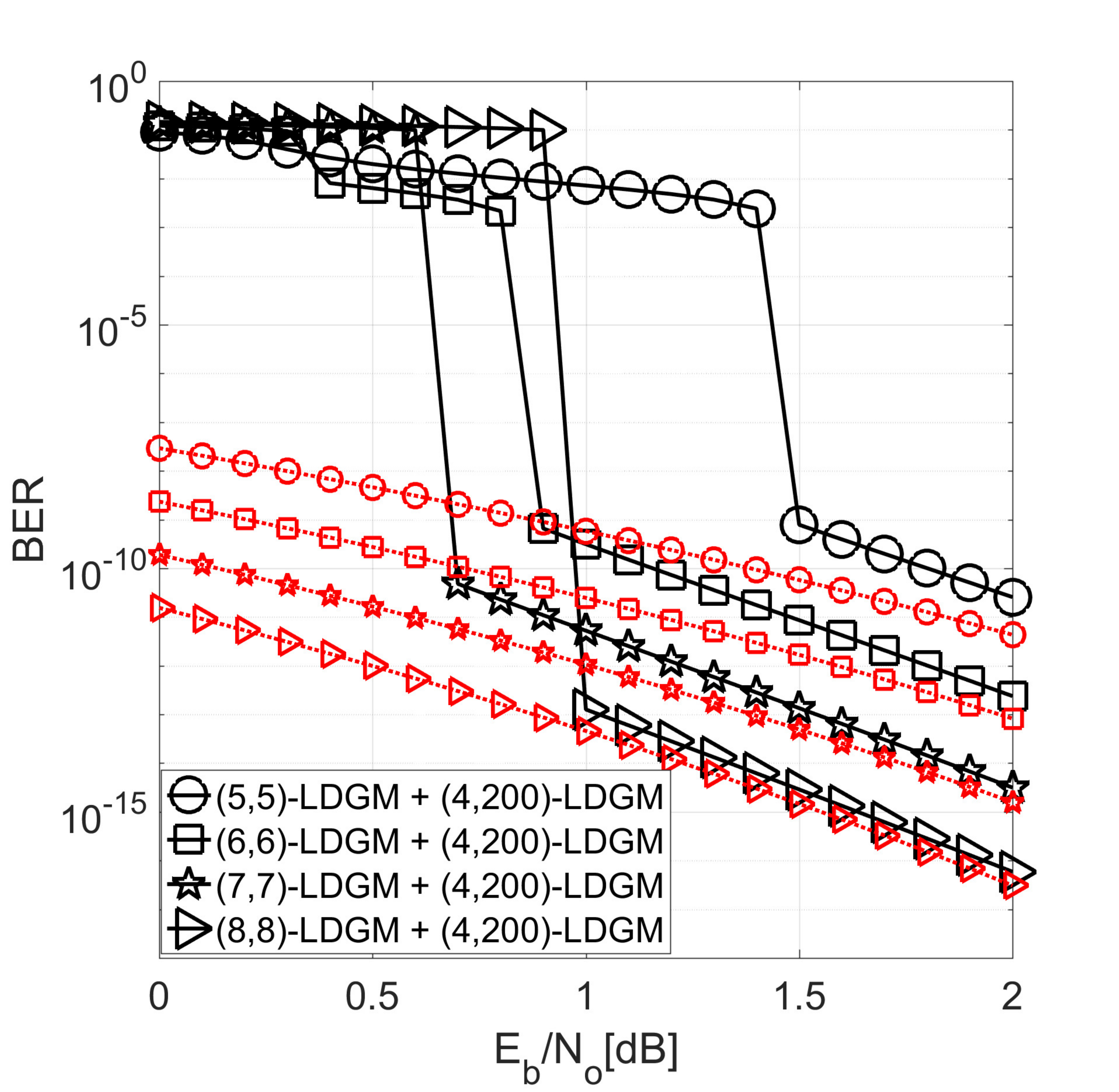}}
\caption{Asymptotic performance (solid lines) and lower bounds (dotted lines) of various SCLDGM codes. Overall code-rate (r) = 0.49.} 
\label{fig_DDE_LDGMhalfRate_LowerBound}
\end{figure}

For the inner LDGM code, we knew that $M_D^{(l)} \leq \left(d_v+1\right) M_{u_0}$. In fact, this $M_D^{(l)}$ is the mean of the input LLRs to the outer decoder through the super-channel. Since $M_D^{(l)} = 2/\sigma_{sup}^2$ and $M_{u_0} = 2/\sigma^2$, we can write the following
\begin{equation}\label{eq_sigma_sup}
\begin{split}
& \frac{2}{\sigma_{sup}^2} \leq (d_v+1) \frac{2}{\sigma^2} \\
& \sigma^2_{sup} \geq \frac{\sigma^2}{(d_v+1)}
\end{split}
\end{equation}

Combining (\ref{eq_E_bound1}) and (\ref{eq_sigma_sup}), we get
\begin{equation}\label{eq_E_boundFinal}
E^{(l)} \geq Q\left( \sqrt{\frac{(d_v^{(o)}+1) (d_v+1) }{\sigma^2}} \right)
\end{equation}

Hence, the decoding error probability of a SCLDGM code under two-step SP decoding is always lower bounded by $Q\left( \sqrt{\left((d_v^{(o)}+1) (d_v+1)\right) / \sigma^2 }\right)$. The exact asymptotic performance of various SCLDGM codes obtained using the two-step DDE along with their respective lower bounds are presented in Fig. \ref{fig_DDE_LDGMhalfRate_LowerBound}. We observe that at higher $E_b/N_o$ values, the exact asymptotic performance of the SCLDGM codes is closely approximated by the lower bound formula while they significantly differ at lower $E_b/N_o$ values. More specifically, we see that at any $E_b/N_o$ value above the decoding threshold, we can just use the lower bound formula to approximately calculate the BER of the code.

\section{Error Floor Eradication using High-Rate LDPC Code as an Outer Code}\label{section_LDGM_LDPC}
\begin{figure}[ht!]
\centering{\includegraphics[width=5.5in,height=2.5in]{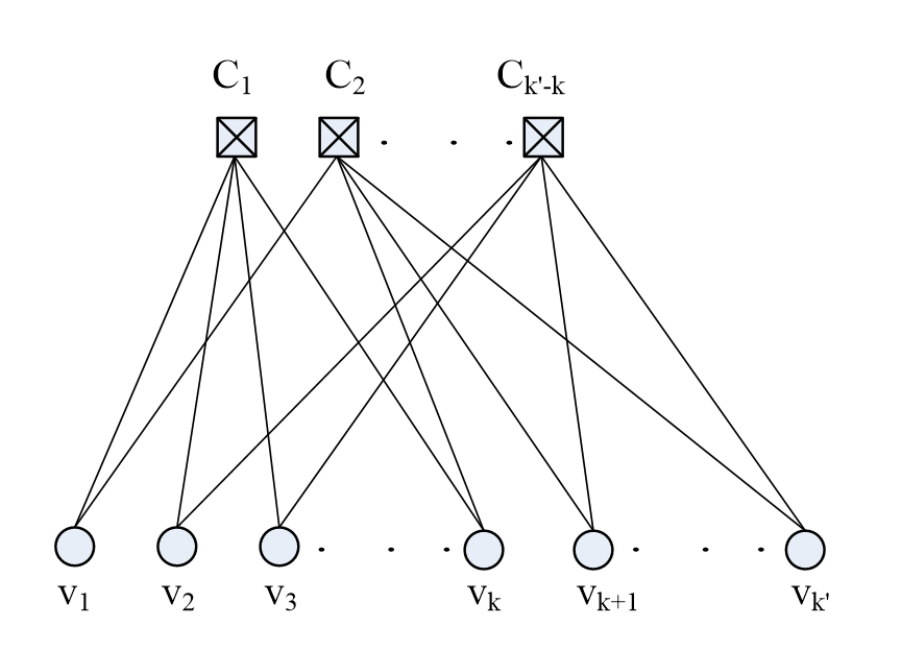}}
\caption{An example of LDPC decoding graph.}
\label{fig_outerDecodingGraphLDPC}
\end{figure}

\begin{figure}[ht!]
\centering{\includegraphics[width=5.5in,height=4in]{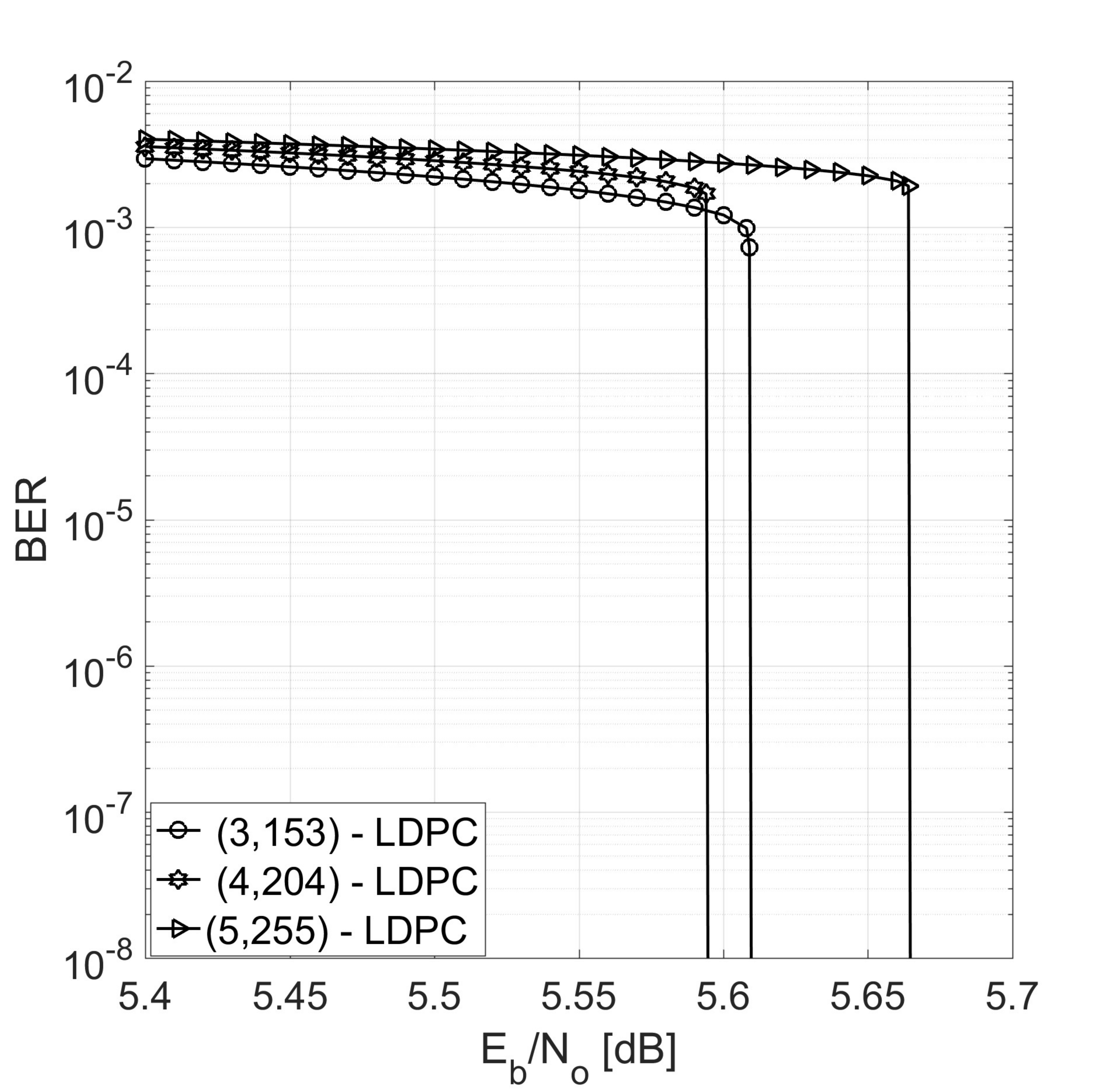}}
\caption{Asymptotic performance of $\left(d_v,d_c\right)$-regular LDPC codes, code-rate = 50/51.}
\label{fig_ldpc50by51Performance}
\end{figure}

\begin{figure}[ht!]
\centering{\includegraphics[width=5.5in, height=4.5in]{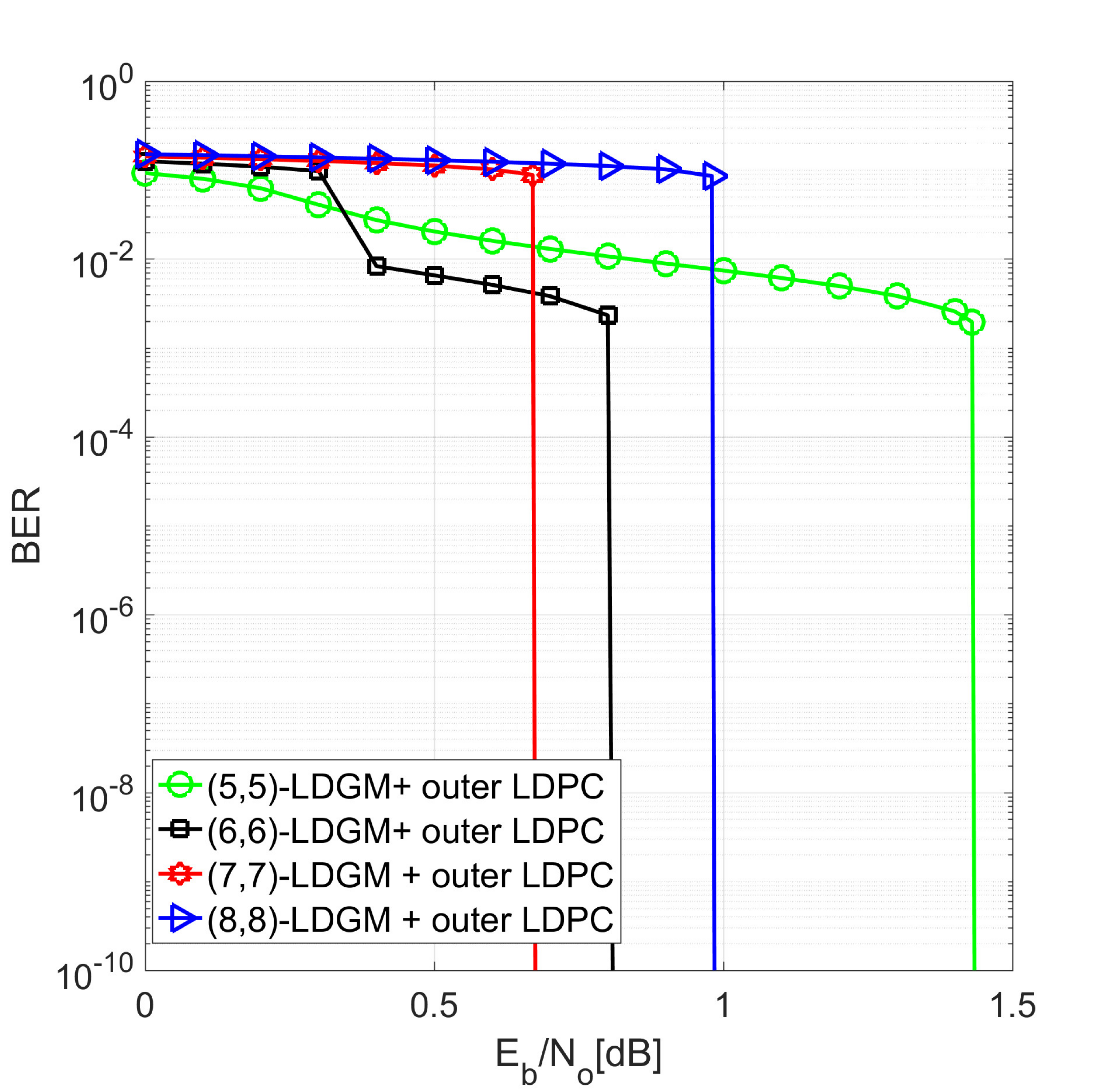}}
\caption{Asymptotic performance of LDPC-GM codes. Inner code used is rate half $(d_v,d_c)$-LDGM code and outer code used is $50/51$ rate (4,204)-LDPC code.  Overall code-rate (r) = 0.49.}
\label{fig_LDGMldpc}
\end{figure}

In the LDPC-GM codes, the high-rate LDPC codes are used as the outer codes. An LDPC code is defined by its parity check matrix $(H)$ which is used to construct its bi-partite decoding graph as shown in Fig. \ref{fig_outerDecodingGraphLDPC}, where $k'$ represent the length of the codeword produced by the LDPC encoding. Let $\bm{w}=[w_1 w_2 \cdots w_{k'}]$ be the LDPC codeword each bit of which are, in fact, represented by VNs $V_1, V_2, \cdots, V_{k'}$ in Fig. \ref{fig_outerDecodingGraphLDPC}. Knowing that in GF(2), $wH^{T}=0$, where $T$ represents the transpose, the CNs $C_1, C_2, \cdots, C_{k'-k}$ are always zero, i.e., in terms of LLR values, the magnitudes are infinite. Hence, unlike in LDGM codes, the messages being passed by LDPC's CNs are independent of any channel estimates since their CNs do not represent any part of the transmitted codeword. They rather continuously evolve as decoding proceeds. Due to this fact, the error floors are absent in the asymptotic curves of the $50/51$-rate regular LDPC code as depicted in Fig. \ref{fig_ldpc50by51Performance}. The decoding threshold of the $50/51$-rate LDPC code and the corresponding critical BER as presented in Table \ref{table_ldpc_thresholds50by51} are almost the same as those of $50/51$-rate LDGM codes presented in Table \ref{table_thresholds50by51}. This suggests that the decoding threshold of the LDPC-GM codes should be as good as the SCLDGM codes, and due to the use of outer LDPC codes, the asymptotic curves of such concatenated codes should be free from the error-floors. As expected, we observed no error-floors for LDPC-GM codes as in Fig. \ref{fig_LDGMldpc}. The outer LDPC code used is of code-rate $50/51$ with $d^{(o)}_v=4$. Hence, the use of high-rate LDPC code as an outer code instead of the high-rate LDGM code helps completely eliminate the error floors without sacrificing the decoding performance.


 \begin{table}[ht]
\centering
\caption{DDE threshold of $50/51$-rate $\left(d_v, d_c \right)$ LDPC code and the corresponding critical BER.}
\label{table_ldpc_thresholds50by51}
\scalebox{1.0}{
\begin{tabular}{|c| c| c| c|}
\hline
$\left(d_v, d_c \right)$ & $(E_b/N_o)_{th}$ & $\sigma^{(o)}_{th}$ & $\mathrm{Q}\left(1/\sigma^{(o)}_{th}\right)$ \\
\hline
$\left(3,153\right)$ & 5.610 & 0.374 & 3.778 $\times 10^{-3}$ \\ 
$\left(4,204\right)$ & 5.595 & 0.375 & 3.831 $\times 10^{-3}$ \\ 
$\left(5,255\right)$ & 5.665 & 0.372 & 3.592 $\times 10^{-3}$ \\ 
\hline
\end{tabular}
}
\end{table}

The two-step DDE method required to obtain the asymptotic curves (Fig. \ref{fig_LDGMldpc}) for the LDPC-GM codes are very similar to the one explained in Section {\ref{section_DDE_SCLDGM}} in the sense that the DDE of the inner LDGM codes is exactly the same while the DDE implementation for the outer code need to consider the LDPC decoding structure. Since, the LDPC-GM codes closely resemble Raptor codes \cite{Raptor}, as in \cite{amritLetter2018, amritCCNC2018}, by using $p^{(l)}_{\bar{D}}\left(\bar{d}\right)$ from the inner decoder as the PMF of the initial LLRs for the outer decoder, the DDE for the outer LDPC code can be easily implemented.

\section{Optimal Code Design}\label{section_Optimization}
DDE-based optimization methods are used in \cite{LDPC_design_Richardson, LDPC_ChungDissertation} to design capacity approaching irregular LDPC codes. Due to the detailed two-step DDE implementation of the SCLDGM codes presented in Section \ref{section_DDE_SCLDGM} and the necessary condition for the successful decoding, it is now possible to use DDE-based optimization approaches to optimally design such concatenated codes. For a given outer code, designing a good SCLDGM code means finding the degree distribution pair $\Lambda(x)$ and $\Omega(x)$ of the inner code such that the overall decoding error probability $E^{(l)}$ given by (\ref{eq_Pe_Raptor_sum}) tends to zero at the lowest possible $E_b/N_o$. This arduous task, however, with the knowledge of the critical BER becomes finding the pair $\Lambda(x)$ and $\Omega(x)$ of the inner code such that $E^{(l)}_{in} \leq \mathrm{Q}\left( 1/\sigma_{th}^{(o)}\right)$ is achieved at the lowest possible $E_b/N_o$. The constraints required for the optimization are (a) $\Lambda(1)=1$, (b) $\Omega(1)=1$, and (c) $E^{(l)}_{in} \leq \mathrm{Q}\left( 1/\sigma_{th}^{(o)}\right)$. The first and second constraints guarantee the valid degree distributions while the third constraint is the necessary condition for the successful decoding.


\begin{figure}[ht!]
\centering{\includegraphics[width=5.5in, height=5in]{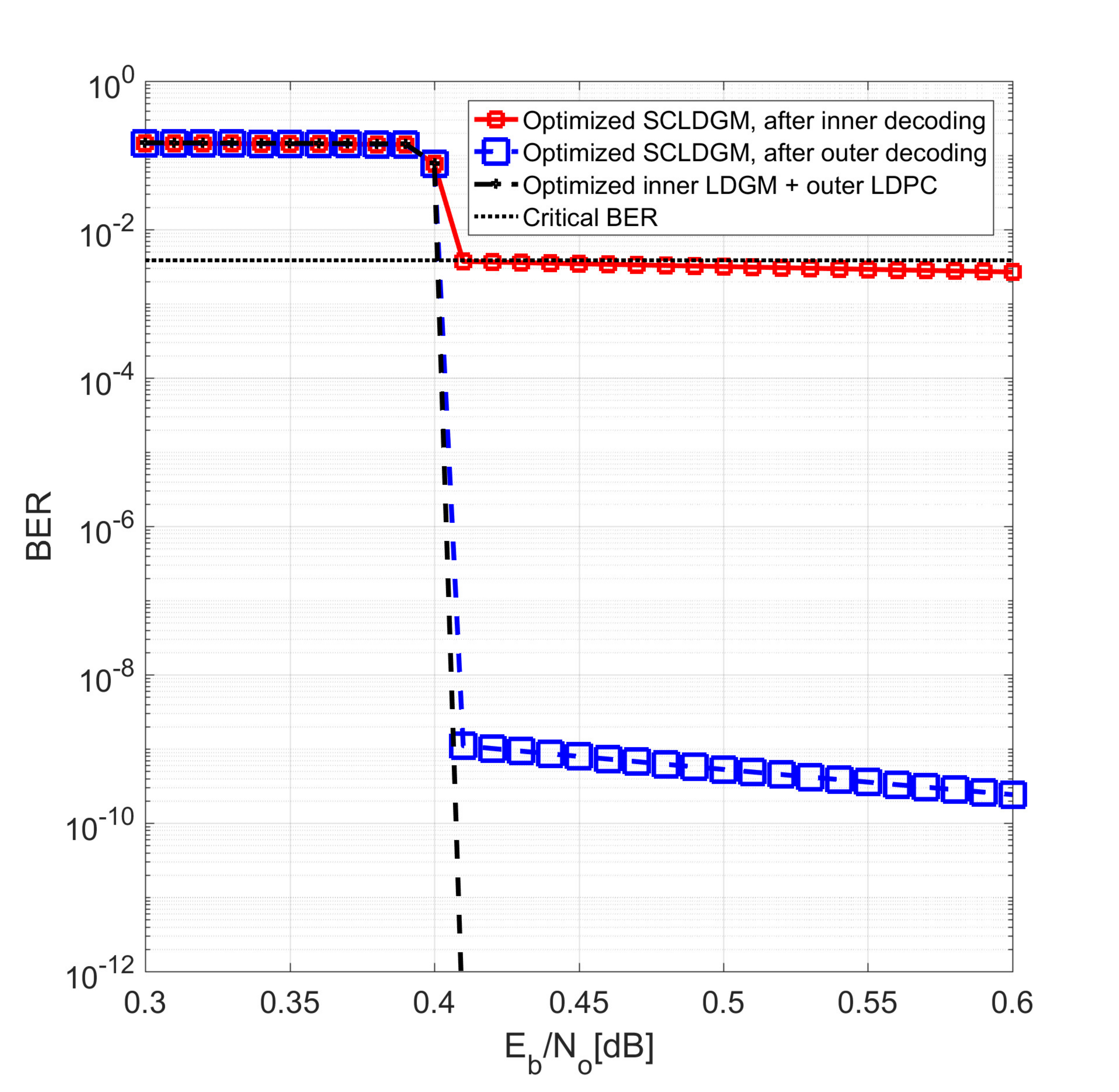}}
\caption{Asymptotic performance of our optimized concatenated codes.  Overall code-rate (r) = 0.49.}
\label{fig_innerOptiVsFullyOptiSCLDGM}
\end{figure}

\begin{figure}[ht!]
\centering{\includegraphics[width=5.5in, height=5in]{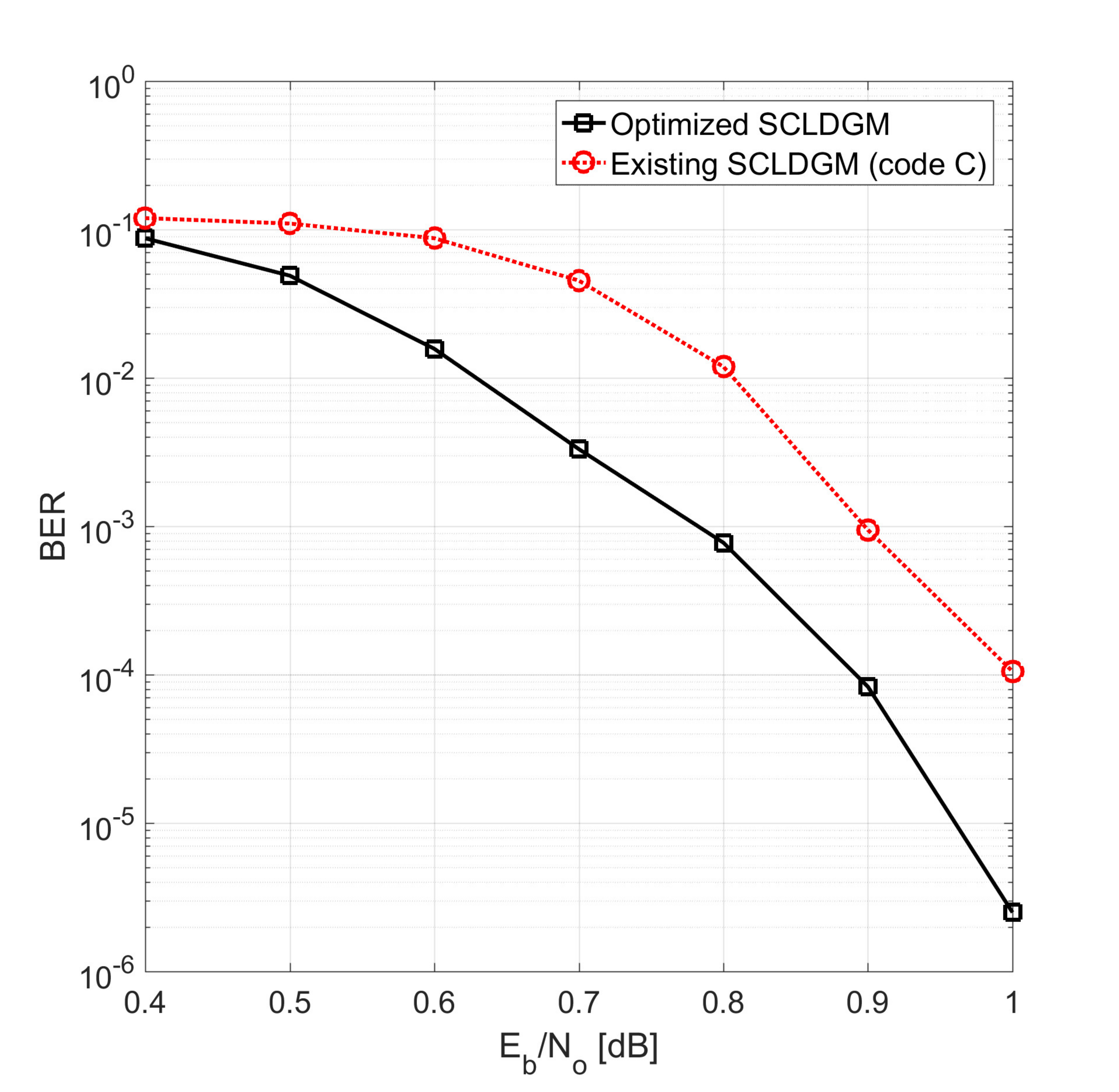}}
\caption{BER performance comparison of our optimized SCLDGM vs. existing SCLDGM code (code C), k=10000.  Overall code-rate (r) = 0.49.}
\label{fig_SCLDGMoptiVsNonOptiBER}
\end{figure}

For the purpose of optimization, we have considered $50/51$ rate $(4,200)$-LDGM code as the outer code. The code-rate $50/51$ is chosen since it is known to be one of the best choices for the outer codes in such concatenated schemes \cite{LDGM_EXIT}. $(4,200)$-LDGM code is used since it is the best regular LDGM code when the code-rate is $50/51$ as shown in Table \ref{table_thresholds50by51}. Hence, for a SCLDGM code with this outer code, the critical BER to be achieved by the inner decoder as known from Table \ref{table_thresholds50by51} is $3.848 \times 10^{-3}$. The optimization process then becomes finding the optimized $\Lambda(x)$ and $\Omega(x)$ for the inner code such that $3.848 \times 10^{-3}$ is achieved at the lowest possible $E_b/N_o$. Actually, we focus on obtaining $\Lambda(x)$ while its pair $\Omega(x)$ is calculated as $\Omega(x)=\Omega_i x^{i} + \Omega_{i+1}x^{i+1}$ for some $i \geq 2$, where the coefficients and the exponents can be easily computed from the knowledge of $\Lambda(x)$ and code-rates \cite{LDPC_ChungDissertation}. We do the minimization of $E^{(l)}_{in}$ starting at slightly higher $E_b/N_o$ and search for degree distributions satisfying the constraints. Once successful, we lower $E_b/N_o$ and repeat the minimization subject to our constraints. Since $\text{Code C} \left((7,7)-\text{LDGM} + (4,200)-\text{LDGM}\right)$ have DDE threshold of $0.68$ dB, the initial $E_b/N_o$ chosen for the optimization process is close to $0.68$ dB, which after each success, is lowered by $0.01$ dB. Due to the monotonicity of the threshold \cite{book_ModetanCodingTheory}, in practice, we speeded up the search process by using the bisection search at a desired level of precision. The lowest $E_b/N_o$ at which such search is successful is called the decoding threshold of the optimized code. The corresponding optimized degree distributions for the inner code we obtained using this process are given by (\ref{eq_inner_opti}). We found that the genetic algorithm based global optimization method called differential evolution \cite{DifferentialEvolution} that has been successfully used to design optimized irregular LDPC codes in \cite{LDPC_design_Richardson} and good erasure codes in \cite{EfficientErasureDiffEvolution} is equally effective in designing optimized SCLDGM codes by incorporating our critical-BER based DDE optimization approach.

\begin{equation}\label{eq_inner_opti}
\begin{split}
\Lambda(x)&= 0.2063x^6 + 0.7472x^7 + 0.0465x^{100}, \\
\Omega(x)&= 0.879x^{11} + 0.121x^{12}.
\end{split}
\end{equation}

The asymptotic performance of our optimized SCLDGM code is depicted in Fig. \ref{fig_innerOptiVsFullyOptiSCLDGM}. The lowest $E_b/N_o$ at which we obtained an inner code that achieved the critical BER during the optimization process was $0.41$ dB. Consequently, we see from the asymptotic curves that at $0.41$ dB and beyond, the overall BER of the SCLDGM code is drastically dropped, making $0.41$ dB the decoding threshold of the optimized SCLDGM code. This code is within $0.26$ dB to the Shannon limit, and clearly outperforms one of the best known SCLDGM codes (i.e, the code C) depicted in Fig. \ref{fig_XXldgm_criticalBER} by a margin of $0.27$ dB. We also observe that the LDPC-GM code that used our optimized inner LDGM code as an inner code provides the same decoding threshold and is free of error floors. In both the schemes, the code-rate of the outer code used is $50/51$ with $d_v^{(o)}=4$. All the results are obtained by using $n_b=10$ and 200 iterations in the DDE. 


The simulation result is presented in Fig. \ref{fig_SCLDGMoptiVsNonOptiBER}. We have used 1000 message blocks each with k=10000 information bits. The maximum number of iterations allowed for the inner and outer decoder is 100. We see that our optimized SCLDGM code outperforms the best SCLDGM code (code C). This also verifies the asymptotic results presented. We would point out that although SCLDGM codes are used in this paper, the concept of the critical BER and the optimization approach can also be exploited in analyzing and designing other serially concatenated error correcting codes.



\section{Conclusion}\label{section_conclusion}
In this paper, we first provided the exact asymptotic performance of the SCLDGM codes over BIAWGN channel using two-step DDE method. We then provided the necessary condition for the successful decoding of these codes. We presented the detailed error-floor analysis and gave lower-bound formulas for the LDGM as well as SCLDGM codes. We showed that the asymptotic performance of the SCLDGM codes beyond the decoding threshold is closely approximated by the lower bound formula. We also provided the asymptotic performance of LDPC-GM codes to demonstrate that by using a high-rate LDPC code as an outer code, the error-floors can be completely removed. Finally, we used the novel critical BER-based DDE optimization method to optimally design SCLDGM codes with improved decoding performance.

\begin{appendix}


For any two LLR messages $L_0$ and $L_1$, based on the product rule, the output LLR $L$ is calculated as $L=2\tanh^{-1}\left(\tanh(L_0/2) \tanh(L_1/2)\right)$, which can also be written as
\begin{equation}\label{eq_productrule_expanded}
\begin{split}
     L & = \log \frac{1+e^{(L_0+L_1)}}{e^{L_0}+e^{L_1}} \\
       & = \log \left(1+e^{(L_0+L_1)}\right) - \log \left(e^{L_0}+e^{L_1}\right).
\end{split}
\end{equation}

For any two real numbers $x$ and $y$, we have
\begin{equation}\label{eq_log_exey}
  \log(e^x+e^y)=\max(x,y) + \log\left(1+e^{-|x-y|}\right).
\end{equation}
Using (\ref{eq_productrule_expanded}) and (\ref{eq_log_exey}), we can write
\begin{equation}\label{eq_u_logexey}
 L = \max(0, L_0+L_1) - \max(L_0,L_1) + \log \frac{1+e^{-|L_0+L_1|}}{1+e^{-|L_0-L_1|}}.
\end{equation}
To prove Lemma 1, it suffices to show $|L|\leq \min(|L_0|,|L_1|)$. Without loss of generality, we assume $|L_0|\leq|L_1|$. Since the output of $\tanh()$ has the same sign with the input of the function, we have $\sign(L)=\sign(L_0)\cdot \sign(L_1)$ as shown in Table \ref{table_sign}.

\begin{table}[ht]
\centering
\caption{Sign of $L$ under product rule}
\label{table_sign}
\scalebox{1}{
\begin{tabular}{|c| c|| c|}
\hline
$\sign(L_0)$ & $\sign(L_1)$ & $\sign(L)$\\
\hline
+ & + & + \\
- & - & + \\
+ & - & - \\
- & + & - \\
\hline
\end{tabular}
}
\end{table}

\begin{itemize}
  \item \textbf{Case I}: $L_0>0$, $L_1>0$, and $|L_0|\leq|L_1|$ \\
  In this case, (\ref{eq_u_logexey}) reduces to
\begin{equation}\label{eq_Case++}
  L = L_0 + \log \frac{1+e^{-|L_0+L_1|}}{1+e^{-|L_0-L_1|}}.
\end{equation}

Under Case I, we can write
\begin{align} \label{eq_caseI_condition}
    |L_0 + L_1| & > |L_0-L_1| \notag \\
     e^{- |L_0 + L_1|} & < e^{- |L_0-L_1|} \notag \\
    \frac{1+e^{- |L_0 + L_1|}}{1+e^{- |L_0-L_1|}} & < 1 \notag \\
    \log \left( \frac{1+e^{- |L_0 + L_1|}}{1+e^{- |L_0-L_1|}} \right) & < 0
\end{align}

Knowing that $L$ must be positive when both $L_0$ and $L_1$ are positive from Table \ref{table_sign}, and using (\ref{eq_Case++}) and (\ref{eq_caseI_condition}), we can write $|L| - |L_0| < 0$, i.e., $|L| < |L_0|$.


  \item \textbf{Case II}: $L_0<0$, $L_1<0$, and $|L_0|\leq|L_1|$ \\
In this case, (\ref{eq_u_logexey}) reduces to
\begin{equation}\label{eq_Case--}
  L = |L_0| + \log \frac{1+e^{-|L_0+L_1|}}{1+e^{-|L_0-L_1|}}.
\end{equation}

Since $|L_0+L_1| > |L_0-L_1|$ holds true as in Case I, $ \log \left( \frac{1+e^{- |L_0 + L_1|}}{1+e^{- |L_0-L_1|}} \right) < 0 $ also holds true. Using this and the knowledge that $L$ must be positive when both $L_0$ and $L_1$ are negative from Table \ref{table_sign}, we can write (\ref{eq_Case--}) as $|L| - |L_0| < 0$, i.e., $|L| < |L_0|$.

  \item \textbf{Case III}: $L_0>0$, $L_1<0$, and $|L_0|\leq|L_1|$ \\
In this case, (\ref{eq_u_logexey}) reduces to
\begin{equation}\label{eq_Case+-}
  L = - |L_0| + \log \frac{1+e^{-|L_0+L_1|}}{1+e^{-|L_0-L_1|}}.
\end{equation}

In this case, $|L_0+L_1| < |L_0-L_1|$. Then, proceeding as in (\ref{eq_caseI_condition}), we get $ \log \left( \frac{1+e^{- |L_0 + L_1|}}{1+e^{- |L_0-L_1|}} \right) > 0 $. Using this and the knowledge that $L$ must be negative when $L_0$ is positive and $L_1$ is negative from Table \ref{table_sign}, we can write (\ref{eq_Case+-}) as $-|L| + |L_0| > 0$, i.e., $|L| < |L_0|$.

  \item \textbf{Case IV}: $L_0<0$, $L_1>0$, and $|L_0|\leq|L_1|$ \\
Similar to Case III, in this case (\ref{eq_u_logexey}) reduces to $L = - |L_0| + \log \frac{1+e^{-|L_0+L_1|}}{1+e^{-|L_0-L_1|}}$, and $ \log \left( \frac{1+e^{- |L_0 + L_1|}}{1+e^{- |L_0-L_1|}} \right) > 0 $ holds true. Since, $L$ must be negative, we can write $-|L| + |L_0| > 0$, i.e., $|L| < |L_0|$.
\end{itemize}

We have shown using these four cases that, $|L| < |L_0|$. Considering the scenario that when both $L_0$ and $L_1$ are zero, $L$ is also zero, we can write $|L| \leq |L_0|$. Since we considered $|L_0| \leq |L_1|$ in all the cases, we can write $|L|\leq \min \left( L_0, L_1\right)$.

By defining $R(a,b) = 2\tanh^{-1}\left(\tanh(a/2) \tanh(b/2)\right)$ for any two real-valued inputs $a$ and $b$, we can write the product rule as $u=R(v_0,R(v_1,R(v_2 \cdots,R(v_{d_c-2},v_{d_c-1}))))$. Since $L=R\left(L_0, L_1 \right)$ and $|L|\leq \min \left( L_0, L_1\right)$, it is straight forward that $|u|\leq \min \left(|v_0|, |v_1|, \cdots |v_{d_c-1}|\right)$.
\end{appendix}

\section*{Acknowledgment}
This work is supported by NASA EPSCoR program under grants NNX13AB31A and NNX14AN38A.


\end{document}